\documentclass[12pt,preprint]{aastex}
\usepackage{natbib}
\usepackage{lscape}
\usepackage[dvips]{color}
\slugcomment{Accepted for publication in the ApJ: December 17, 2007}

\newcommand       \msun        	{$M_{\odot}$}

\newcommand	     \cc             {cm$^{-3}$}

\newcommand       \mic        	 {$\mu$m}

\newcommand 	      \kms            {km~s$^{-1}$}

\newcommand		\spitzer	 	{{\it Spitzer}}
\newcommand		\chandra 	{{\it Chandra}}
\newcommand		\hst 	{{\it HST}}
\newcommand		\irx 	{$IRX$}

\begin{document}

\title{Infrared and X-Ray Evidence for Circumstellar Grain Destruction by the Blast Wave of Supernova~1987A}

\author{Eli Dwek\altaffilmark{1}, Richard G. Arendt \altaffilmark{2}, Patrice Bouchet\altaffilmark{3}, David N. Burrows\altaffilmark{4}, Peter Challis \altaffilmark{5}, I. John Danziger\altaffilmark{6},    James M.~De
Buizer\altaffilmark{7}, Robert D. Gehrz\altaffilmark{8}, Robert P. Kirshner\altaffilmark{5}, Richard McCray\altaffilmark{9}, Sangwook Park\altaffilmark{4}, Elisha F. Polomski\altaffilmark{8}, and Charles E. Woodward\altaffilmark{8}}

\altaffiltext{1}{Observational Cosmology Lab., Code 665; NASA Goddard
Space Flight Center, Greenbelt, MD 20771, U.S.A., e-mail: eli.dwek@nasa.gov}
\altaffiltext{2}{Science Systems \& Applications, Inc. (SSAI), Code 665, NASA Goddard Space Flight Center, Greenbelt MD, 20771, U.S.A.}  
\altaffiltext{3}{DSM/DAPNIA/Service d'Astrophysique, CEA/Saclay, F-91191 Gif-sur-Yvette; Patrice.Bouchet@cea.fr}
\altaffiltext{4}{Department of Astronomy and Astrophysics, Pennsylvania State University, 525 Davey
Laboratory, University Park, PA 16802, U.S.A.} 
\altaffiltext{5}{Harvard-Smithsonian, CfA, 60 Garden St., MS-19, Cambridge, MA 02138, U.S.A.}
\altaffiltext{6}{Osservatorio Astronomico di Trieste, Via Tiepolo, 11,
Trieste, Italy} 
\altaffiltext{7}{Gemini Observatory, Southern
Operations Center, c/o AURA, Casilla 603, La Serena, Chile}
\altaffiltext{8}{Department of Astronomy, University of Minnesota, 116 Church St., SE, Minneapolis, MN 55455, U.S.A.}
\altaffiltext{9}{Joint Inst. Lab. Astrophysics, University of Colorado, Boulder, CO 80309-0440, U.S.A.}

\begin{abstract}
Multiwavelength observations of supernova remnant (SNR)~1987A show that its morphology and luminosity are rapidly changing at X-ray, optical, infrared, and radio wavelengths as the blast wave from the explosion expands into the circumstellar equatorial ring, produced by mass loss from the progenitor star.    
The observed infrared (IR) radiation arises from the interaction of dust grains that formed in mass outflow with the soft X-ray emitting plasma component of the shocked gas. {\it Spitzer} IRS spectra at 5 - 30~\mic\ taken on day 6190 since the explosion show that the emission arises from $\sim 1.1\times 10^{-6}$~\msun\ of silicate grains radiating at a temperature of $\sim 180\pm^{20}_{15}$~K. Subsequent observations on day 7137 show that the IR flux had increased by a factor of 2 while maintaining an almost identical spectral shape. 
The observed IR-to-X-ray flux ratio (\irx) is consistent with that of a dusty plasma with standard Large Magellanic Cloud dust abundances. \irx\ has decreased by a factor of $\sim 2$ between days 6190 and 7137, providing the first direct observation of the ongoing destruction of dust in an expanding SN blast wave on dynamic time scales. Detailed models consistent with the observed dust temperature, the ionization timescale of the soft X-ray emission component, and the evolution of \irx\ suggest that the radiating silicate grains are immersed in a $3.5\times 10^6$~K plasma with a density of $(0.3-1)\times10^4$~\cc, and have a size distribution that is confined to a narrow range of radii between 0.023 and 0.22~\mic. Smaller grains may have been evaporated by the initial UV flash from the supernova.        

\end{abstract}

\section{INTRODUCTION}
On February 23, 1987, Supernova 1987A (SN~1987A), the brightest supernova since Kepler's SN in 1604, exploded in the Large Magellanic Cloud (LMC). About ten years thereafter, the energy output from the supernova became dominated by the interaction of its blast wave with the inner equatorial ring (ER), a dense ring of gas located at a distance of about 0.7 lyr from the center of the explosion, believed to be produced by mass loss from the progenitor star. The ER is being repeatedly observed at optical wavelengths with the {\it Hubble Space Telescope} ({\it HST}) \citep{pun02}, at X-ray energies with the {\it Chandra X-ray Observatory} \citep{park06a, park07}, at radio frequencies with the Australian Telescope
Compact Array (ATCA) \citep{manchester05}, and in the mid-IR with the Gemini South observatory \citep{bouchet04, bouchet06} and the {\it Spitzer} observatory \citep{bouchet04, dwek07a}. The morphological changes in its appearance in these different wavelength regimes (presented in Figure 6 of McCray 2007) reveal the progressive interaction of the  SN blast wave with the ER. The interaction regions appear as hot spots in the {\it HST} images, representing the shocked regions of finger-like protrusions that were generated by Rayleigh-Taylor instabilities in the interaction of the wind from the progenitor star with the ER (see Figure 2 in McCray 2007). 

The X-ray emission is thermal emission from the very hot plasma, and consists of two main characteristic components: a hard ($kT \approx 2$~keV) component representing a fast shock propagating into a low density medium, and a soft ($kT \approx 0.3$~keV) component representing a decelerated shock propagating into the denser protrusions \citep{park07, zhekov06}. The optical emission arises from the gas that is shocked by the blast wave transmitted
through the dense protrusions in the ER \citep{pun02}, and the radio emission is synchrotron radiation from
 electrons accelerated by the reverse shock \citep{manchester05}. The mid-IR emission spectrum is comprised of line and continuum emission.
The lines most probably originate from the optically bright dense knots. The continuum that
dominates the spectrum is thermal emission from dust  that was formed in the post main sequence wind of the progenitor star before it exploded. This dust could either be located in the shocked X-ray emitting  plasma and heated by electronic collisions or in the optical  knots and radiatively heated by the shocks giving rise to the optical emission \citep{polomski04, bouchet06}. Dust has also formed in the ejecta of the supernova about 530 days after the explosion, as evidenced by optical and IR observations of SN1987A \citep{lucy91, gehrz90, moseley89b, wooden93}, but its current contribution to the total mid-IR emission is negligible \citep{bouchet04}. 

Because the Gemini 11.7~\mic\ image correlates well with both the X-ray and optical emission, the possibility that the dust is radiatively heated in the  knots was considered in detail by \cite{bouchet06}. Estimated dust temperatures of $\sim  125$~K fell short of the observed value of $\sim 180$~K but given the uncertainties in the model parameters, this discrepancy could not firmly  rule out this possible scenario for the location and heating mechanism  of the dust. However, the combined IR and X-ray observations can be more readily explained if the dust resides in the shocked regions of the finger-like protrusions that give rise to the observed soft X-ray emission \citep{zhekov06}. We will adopt this scenario as our working hypothesis, and derive a self-consistent model for the composition, abundance, and size distribution of the dust to explain the evolutionary changes in the IR and X-ray fluxes resulting from the collisional heating and the destruction of the dust grains by the ambient plasma.

We first review the basic physical principles that determine the temperature of collisionally heated dust, and describe how the IR emission can be used to probe the physical condition of the X-ray emitting  plasma. We also discuss what information can be derived from the comparison of the IR and X-ray fluxes from the gas (\S2.1). The physics of dust particles in a hot gas is discussed in more detail by Dwek (1987) and Dwek \& Arendt (1992). In \S3 we present a simple analytical model for the evolution of the grain size distribution and total dust mass in the gas that is swept up by an expanding SN blast wave. In \S4 we present {\it Spitzer} low resolution $5-30$~\mic\ IRS spectra obtained on days 6190 and 7137 after the explosion. The IR spectra are used to derive the temperature and composition of the shock heated dust. We use IR and X-ray observations of the SN to constrain the grain size distribution and the time at which the SN blast wave first crashed into its dusty surroundings. The results of our paper are summarized in \S5. 

\section{The Infrared Diagnostics of a Dusty X-ray Plasma}

The morphological similarity between the X-ray and mid-IR images of SN~1987A suggests that  the IR emission arises from dust that is collisionally heated by the X-ray emitting gas. Simple arguments presented below show that, under certain conditions, the IR luminosity and spectrum of a dusty plasma can be used as a diagnostic for the physical conditions of the gas and the details of the gas-grain interactions. Details of the arguments can be found in \cite{dwek87c} and \cite{dwek92a}. 
 
\subsection{The Dust Temperature as a Diagnostic of Electron Density}
The collisional heating rate, ${\cal H}$(erg~s$^{-1}$), of a dust grain of radius $a$ embedded in a hot plasma is given by:
\begin{equation}
\label{hcoll}
{\cal H} = \pi a^2\ \sum_j n_j\ v_j\ {\cal E}_j
\end{equation}
where $n_j$ is the number density of the $j$-th plasma constituent,  $v_j$, its thermal velocity, and ${\cal E}_j$ its thermally-averaged energy deposition in the dust. In all the following, we will assume that the ion and electron temperatures are equal. Then $v_e \gg v_{ion}$, and the dust heating rate is dominated by electronic collisions. 

Let $E_{dep}$ be the thermally-averaged energy deposited by electrons in the solid. If most electrons are stopped in the dust then $E_{dep}$ is, on average, equal to the thermal energy of the electrons, that is, $E_{dep} \propto T_e$, where $T_e$ is the electron temperature. On the other hand, if most incident electrons go entirely through the grains, then $E_{dep}$ is proportional to the electron stopping power in the solid, defined as $\rho^{-1}(dE/dx)$. At the energies of interest here the electronic stopping power has an energy dependence of $(dE/dx) \sim E^{-1/2}$  \citep{iskef83}, or $dE \sim E^{-1/2}\, dx$ so that $E_{dep} \sim T_e^{-1/2}\, a$.  

The functional dependence of the grain heating rate on gas density and temperature is then given by:
\begin{eqnarray}
{\cal H}  & \sim &   a^2\ n_e\ T_e^{3/2}  \qquad \ \ \ {\rm when\ electrons\ are\ stopped\ in\ the\ grain} \\ \nonumber
 & \sim &  a^3\ n_e \qquad \qquad \ \ \ {\rm when\ electrons\ go\ through\ the\ grain}
\end{eqnarray}
where we used the fact that $v_e \sim T_e^{1/2}$. 

The radiative cooling rate, ${\cal L}$(erg~s$^{-1}$), of the dust grain with temperature $T_d$ by IR emission is given by:
\begin{eqnarray}
\label{lrad}
{\cal L} & = & \pi a^2\ \sigma T_d^4\ \left< Q \right> \\ \nonumber
& \sim & \pi a^3\ \sigma T_d^{4+\beta}
\end{eqnarray}
where $\sigma$ is the Stefan-Boltzmann constant, and $\left< Q \right> \propto a\, T_d^{\beta}$ is the Planck-averaged value of the dust emissivity, $Q(\lambda) \propto \lambda^{-\beta}$, where the value of emissivity index, $\beta$, is $\approx 1-2$.

In equilibrium, ${\cal H} = {\cal L}$, and the dust temperature dependence on plasma density and temperature can be written as:
\begin{eqnarray}
\label{tdust}
T_d & \sim & \left({n_e \over a}\right)^{\gamma}\, T_e^{3\gamma/2} \qquad \ \ \ {\rm when\ electrons\ are\ stopped\ in\ the\ grain} \\ \nonumber
 & \sim & n_e^{\gamma} \qquad  \qquad \ \ \ \qquad {\rm when\ electrons\ go\ through\ the\ grain} 
\end{eqnarray}
where $\gamma \equiv 1/(4+\beta)$. These simple arguments show that when the gas temperature is sufficiently high, and the grain size sufficiently small so that most electrons go through the grain, the dust temperature depends only on the plasma density. 

\begin{figure}[ht]
 \begin{tabular}{cc}
  \includegraphics[width=3.0in]{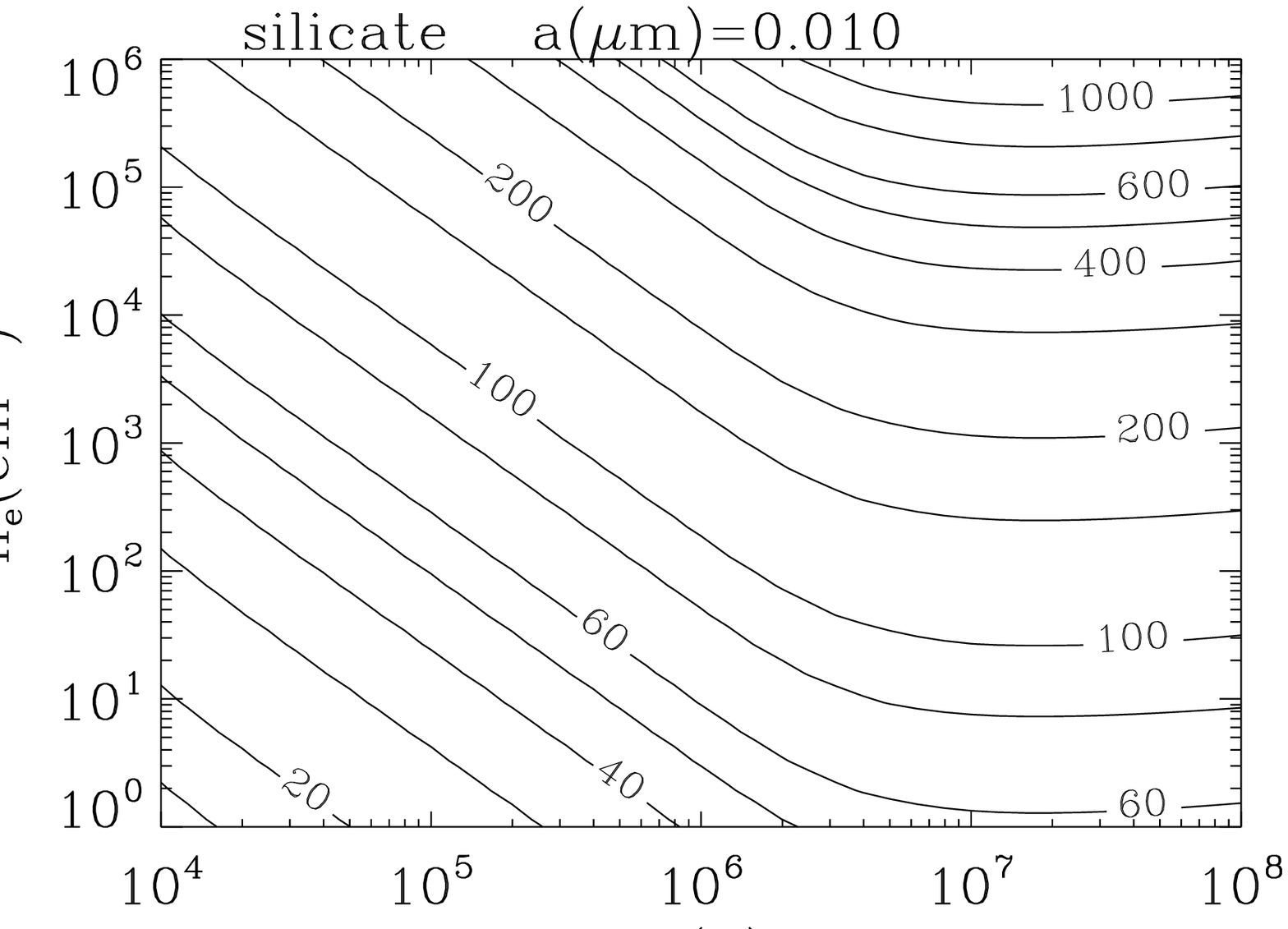} &
    \includegraphics[width=3.0in]{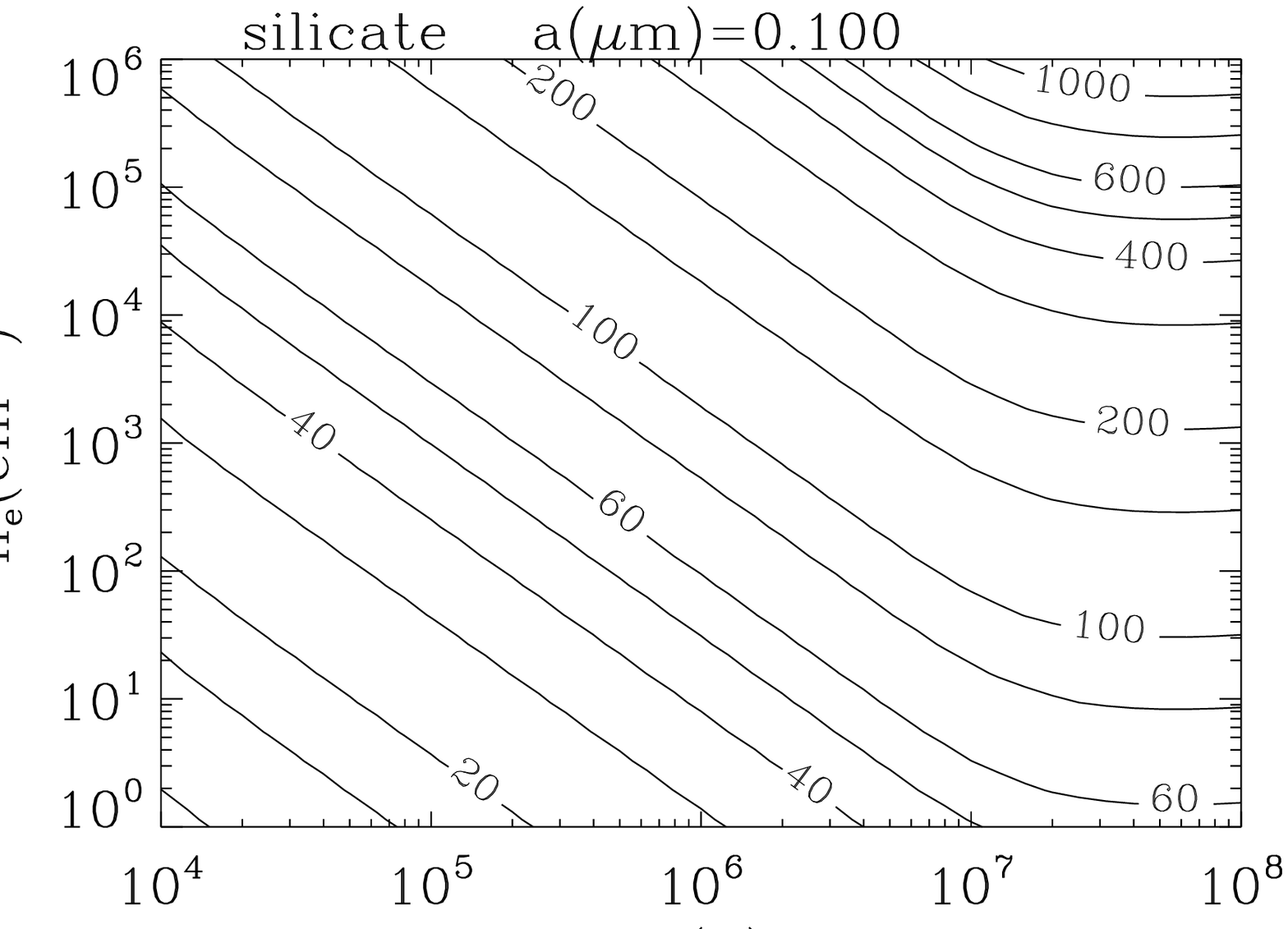}
  \end{tabular} 
  \caption{{\footnotesize Contour plot of the equilibrium temperature of 0.01~\mic\ (left) and 0.10~\mic\ (right) silicate grains as a function of electron density and temperature. Above temperatures of $\sim 5\times10^6$~K ($\sim 3\times10^7$~K) the 0.01~\mic\ (0.1~\mic) grains become transparent to the incident electrons, and the dust temperature is only a function of electron density. At lower temperatures, different combination of plasma density and temperature can heat the dust to the same temperature.  We note here that a similar figure (Figure 15 in Bouchet et al. 2006) was mislabeled, and actually corresponds to contour levels of silicate dust temperature for grain radius of $a=0.0030$~\mic.}}
  \label{tempeq}
\end{figure}

Figure \ref{tempeq} depicts contour levels of the dust temperature as a function of electron density and temperature for 0.01~\mic\ and 0.10~\mic\ silicate grains. The figure shows that above a certain gas temperature, its value depending on the grain radius, most of the electrons go through the grain and the dust temperature is essentially determined by the electron density. Under these conditions, the IR spectrum from the collisionally-heated grains becomes an excellent diagnostic of the density of the X-ray emitting gas. However, when the grains are large enough to stop the electrons, the dust temperature only constrains the possible combination of plasma temperature and density.

\begin{figure}
 \begin{tabular}{cc}
  \includegraphics[width=3.0in]{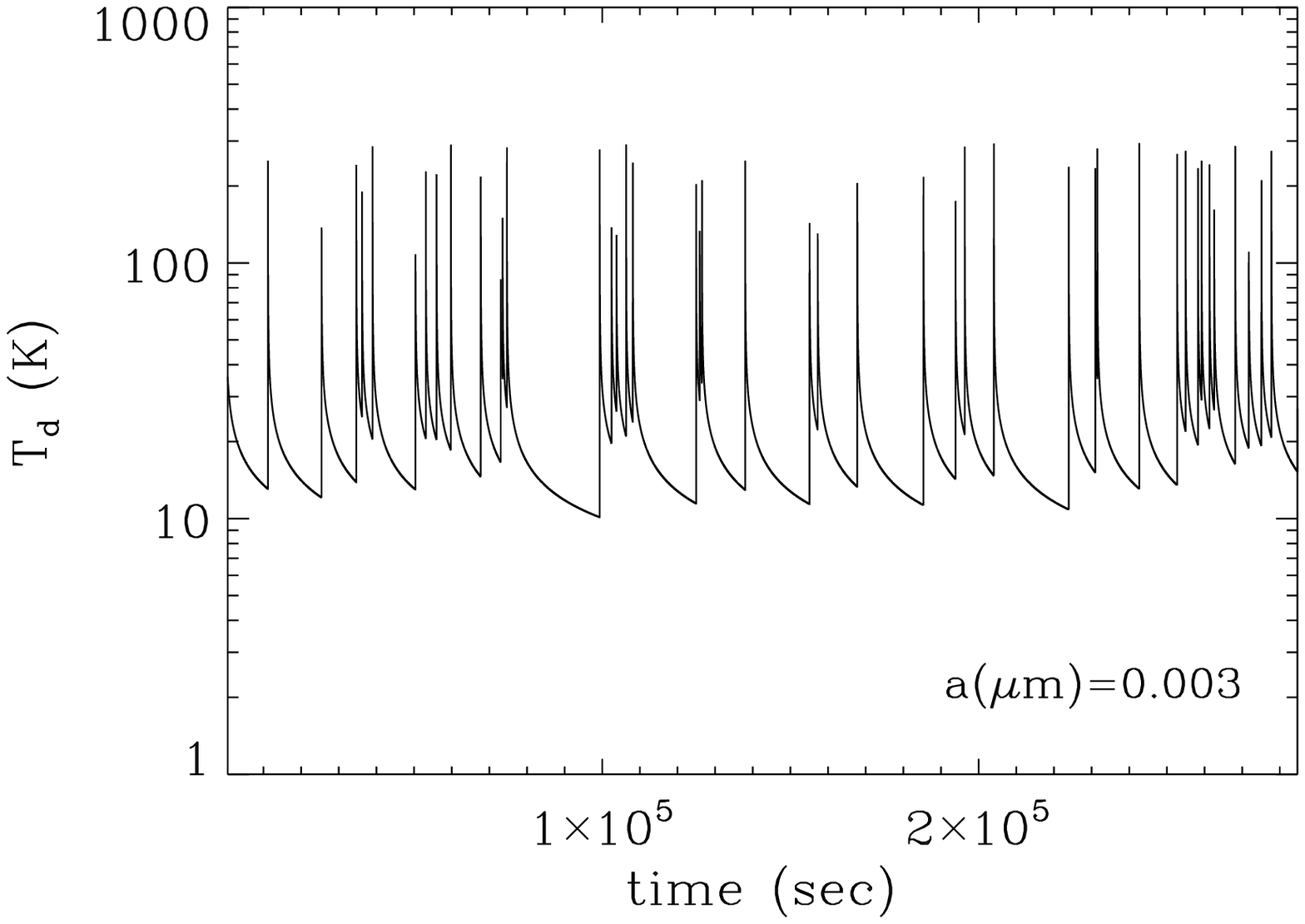} &
  \includegraphics[width=3.0in]{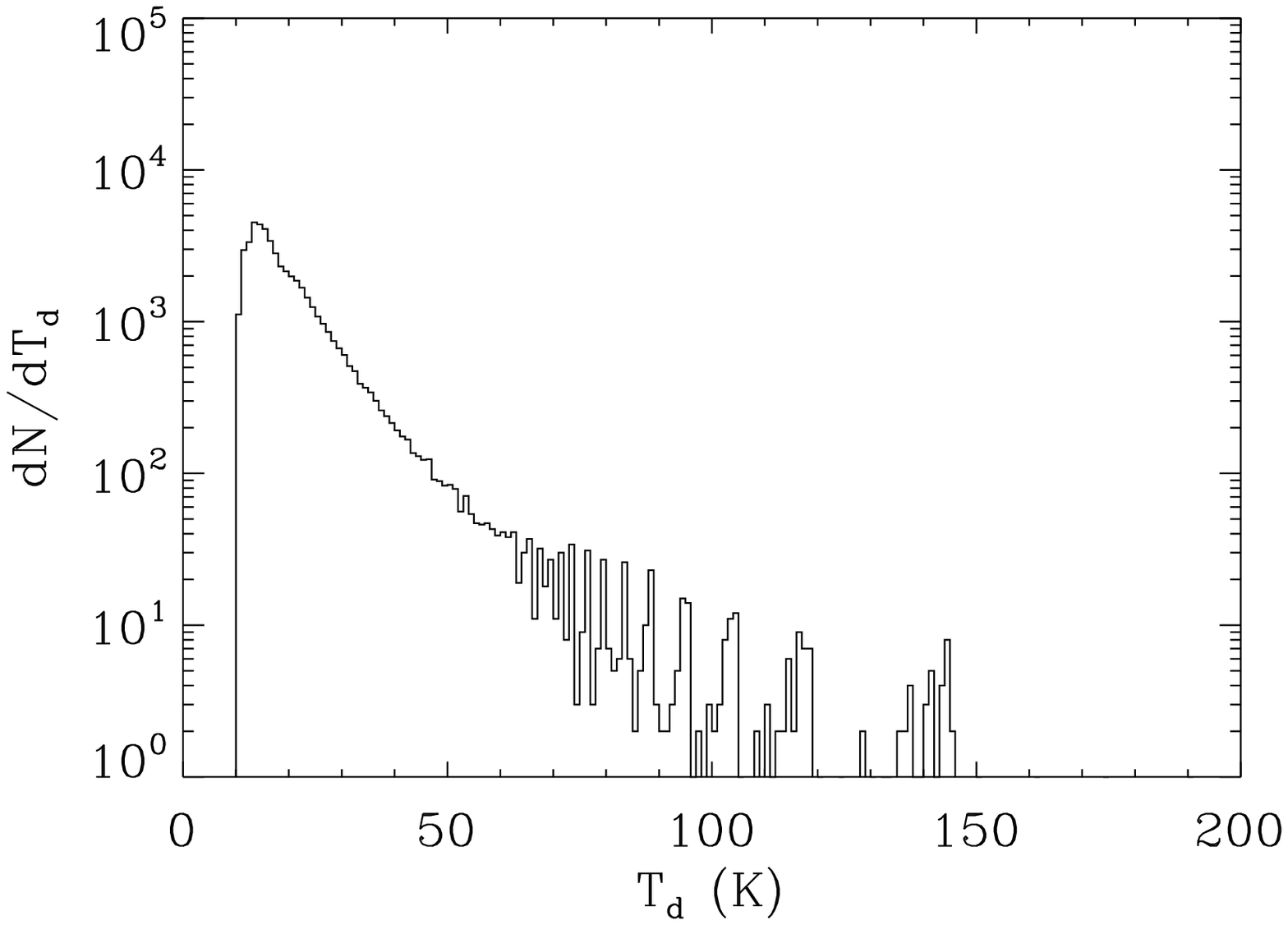}\\
  \includegraphics[width=3.0in]{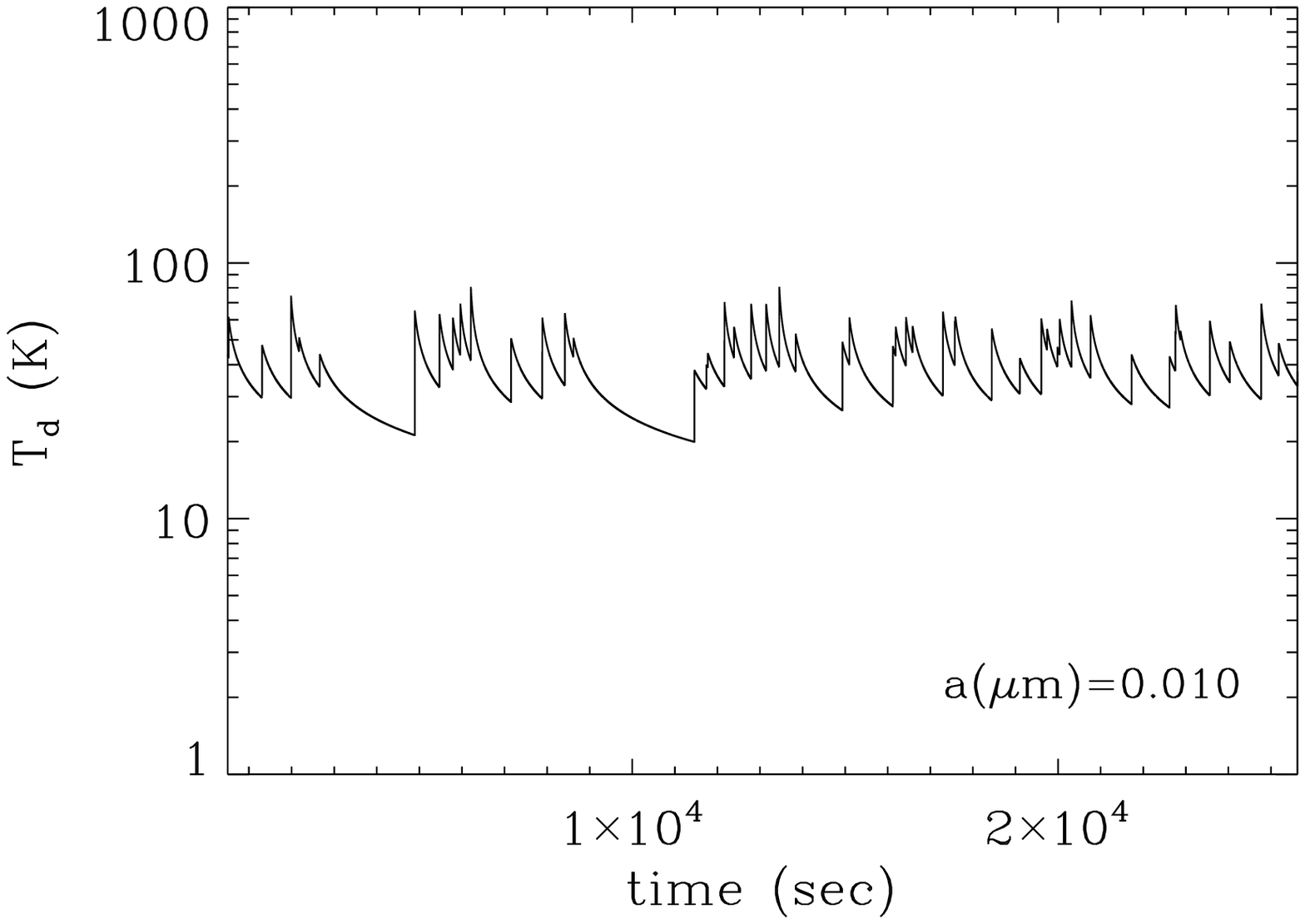} &
  \includegraphics[width=3.0in]{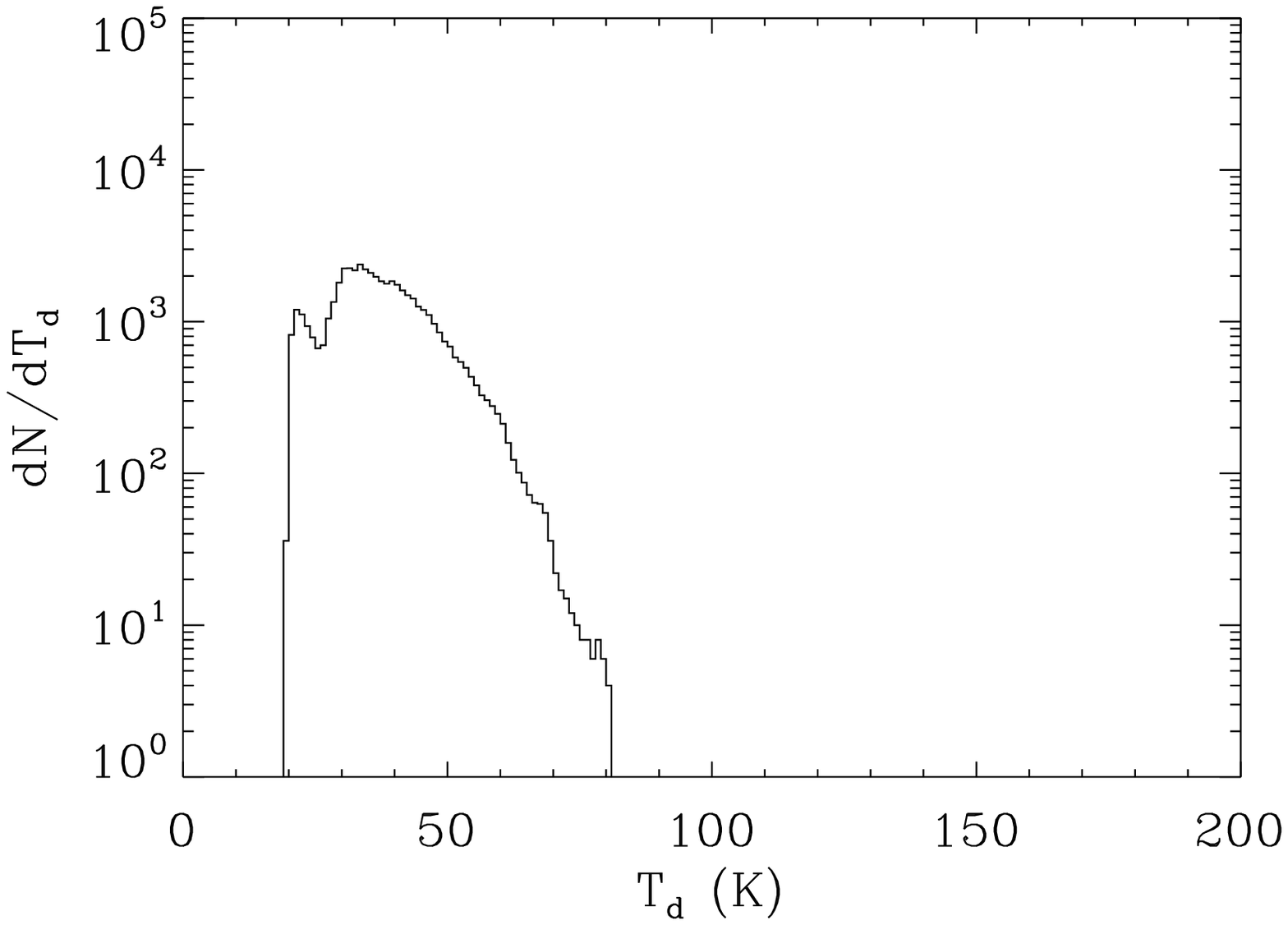}\\
  \includegraphics[width=3.0in]{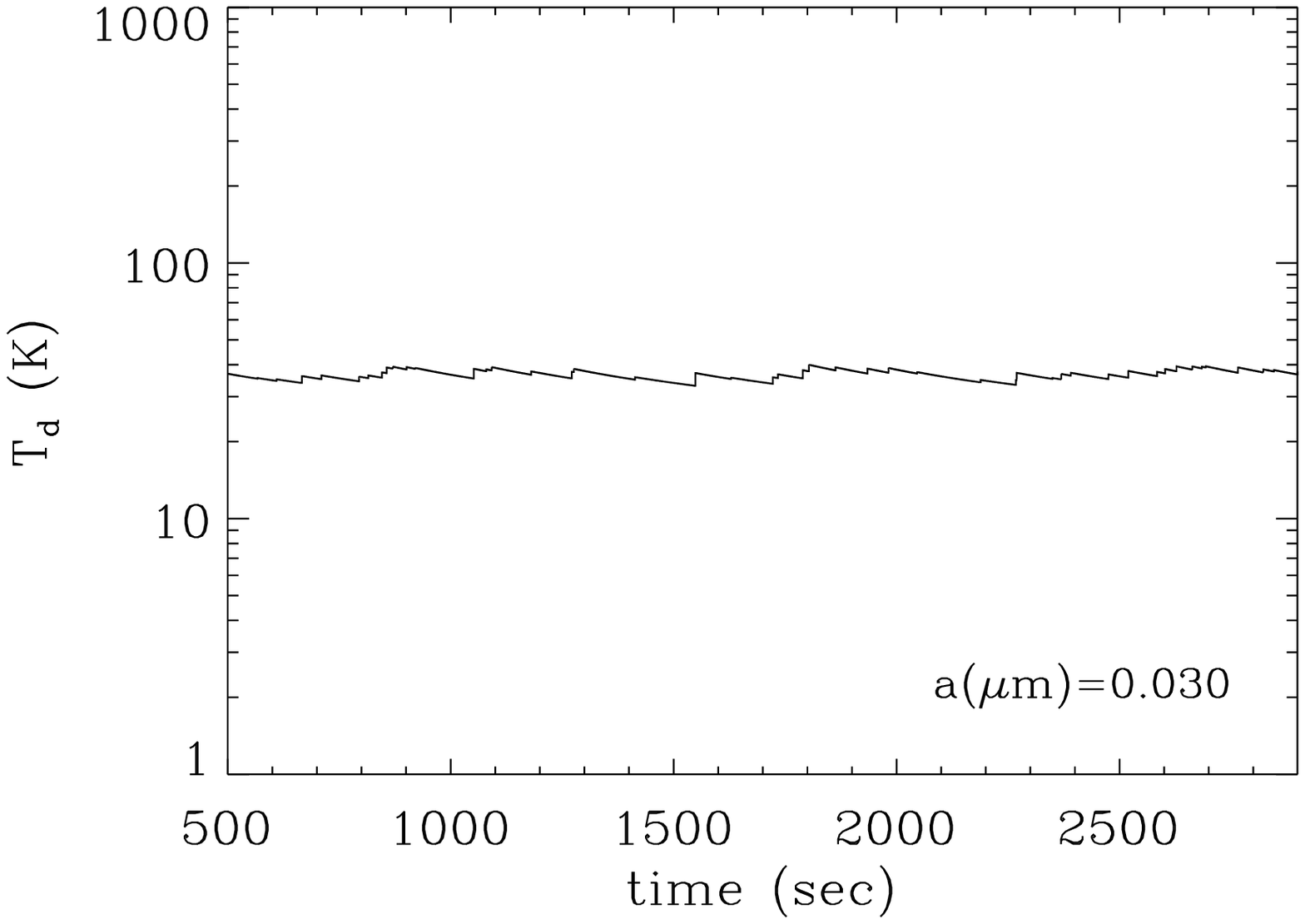} &
  \includegraphics[width=3.0in]{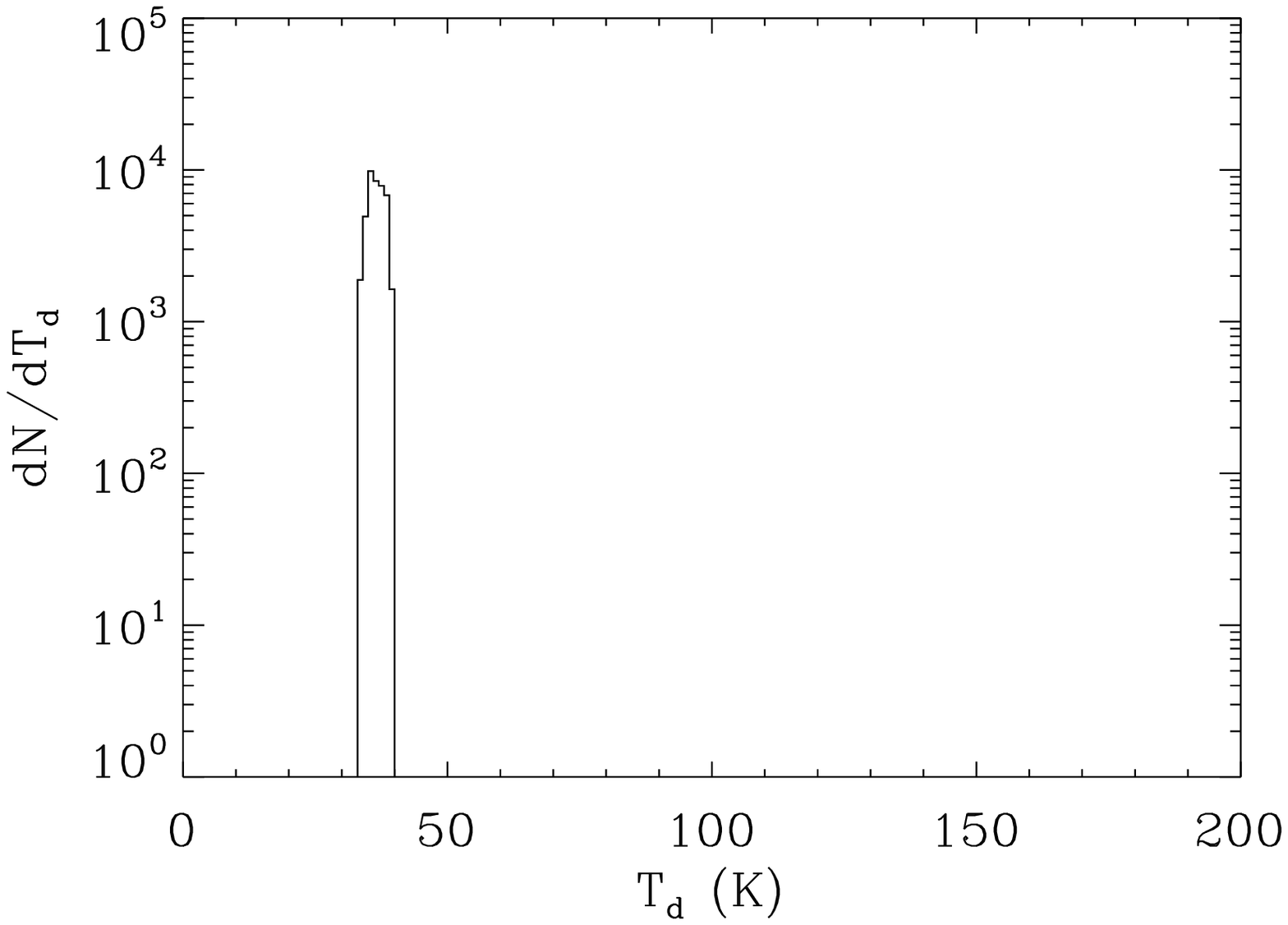} \\
\end{tabular}
\caption{{\footnotesize The stochastic heating of silicate grains in a hot X-ray emitting gas characterized by a temperature of $T_g=10^6$~K, and electron density $n_e=1$~\cc\ for dust grains of different radii. {\bf Left column:} The temperature fluctuations as a function of time. {\bf Right column:} The histogram of the fluctuations. As the grain size increases, the fluctuations get smaller, and the probability distribution of dust temperatures becomes strongly peaked around the equilibrium temperature of $\sim 38$~K. }}
\label{tfluc}
\end{figure}

\subsection{The Stochastic Heating of Grains by Electronic Collisions}
When dust grains are sufficiently small, a single electronic collision can deposit an amount of energy in the dust that is significantly larger that its enthalpy, causing a surge in dust temperature. If additionally, the time interval between successive electronic collisions is larger than the dust cooling time, the grain temperature will be fluctuating with time \citep{dwek87c, dwek92a}.  Figure \ref{tfluc} depicts a simulation of the stochastic heating of 0.003, 0.01, and 0.03~\mic\ silicate grains immersed in a hot X-ray emitting gas characterized a temperature $T_g = 10^6$~K, and an electron density $n_e=1$~\cc. The left column shows the temperature fluctuations as a function of time, and the right column the histogram of the grain temperature. As the grain size increases, the fluctuations get smaller, and the histogram becomes strongly peaked around the equilibrium dust temperature of $\sim 38$~K, in  this example.

\begin{figure}[ht]
\begin{tabular}{cc}
  \includegraphics[width=3.0in]{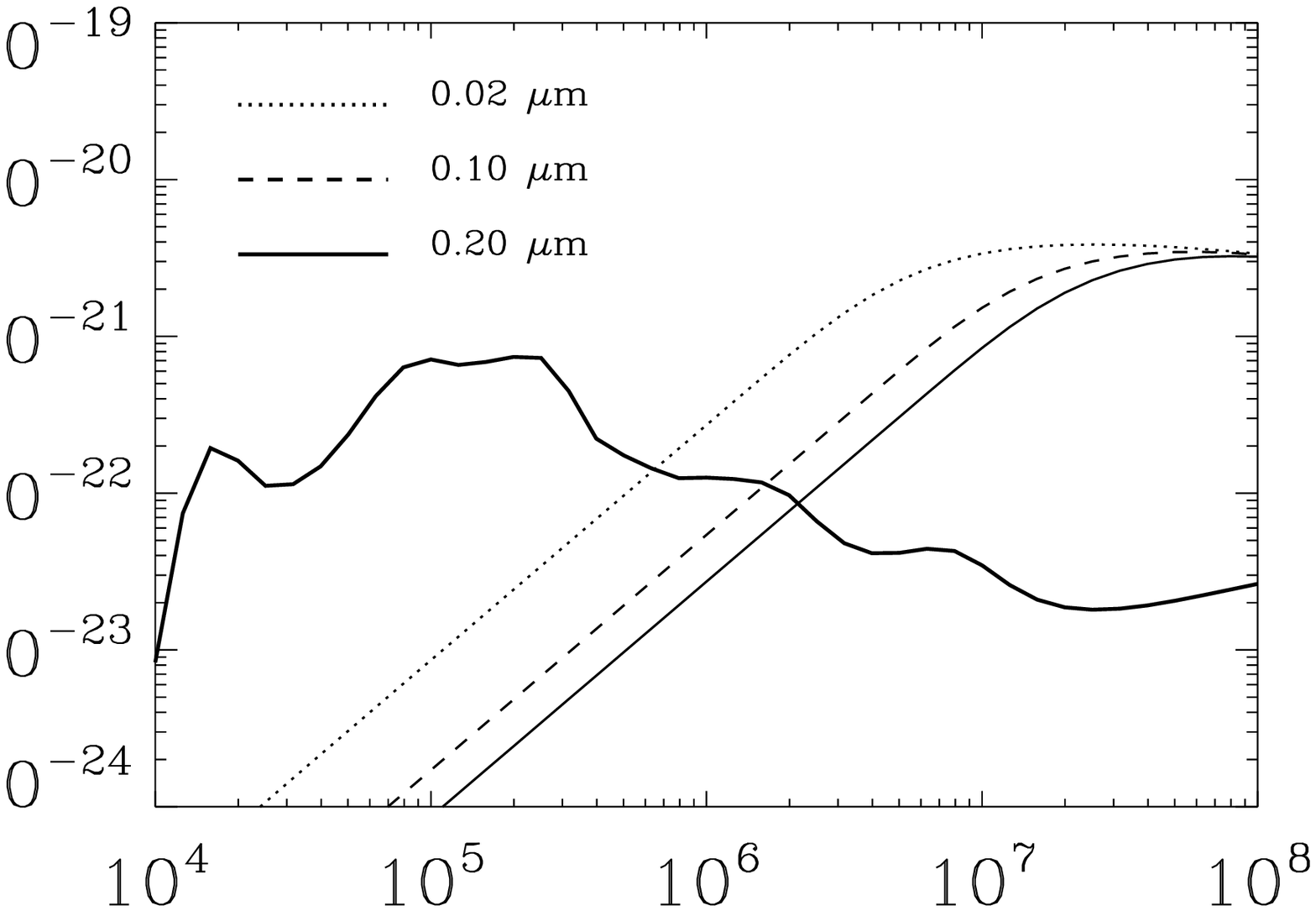} &
  \hspace{0.3in} 
   \includegraphics[width=3.0in]{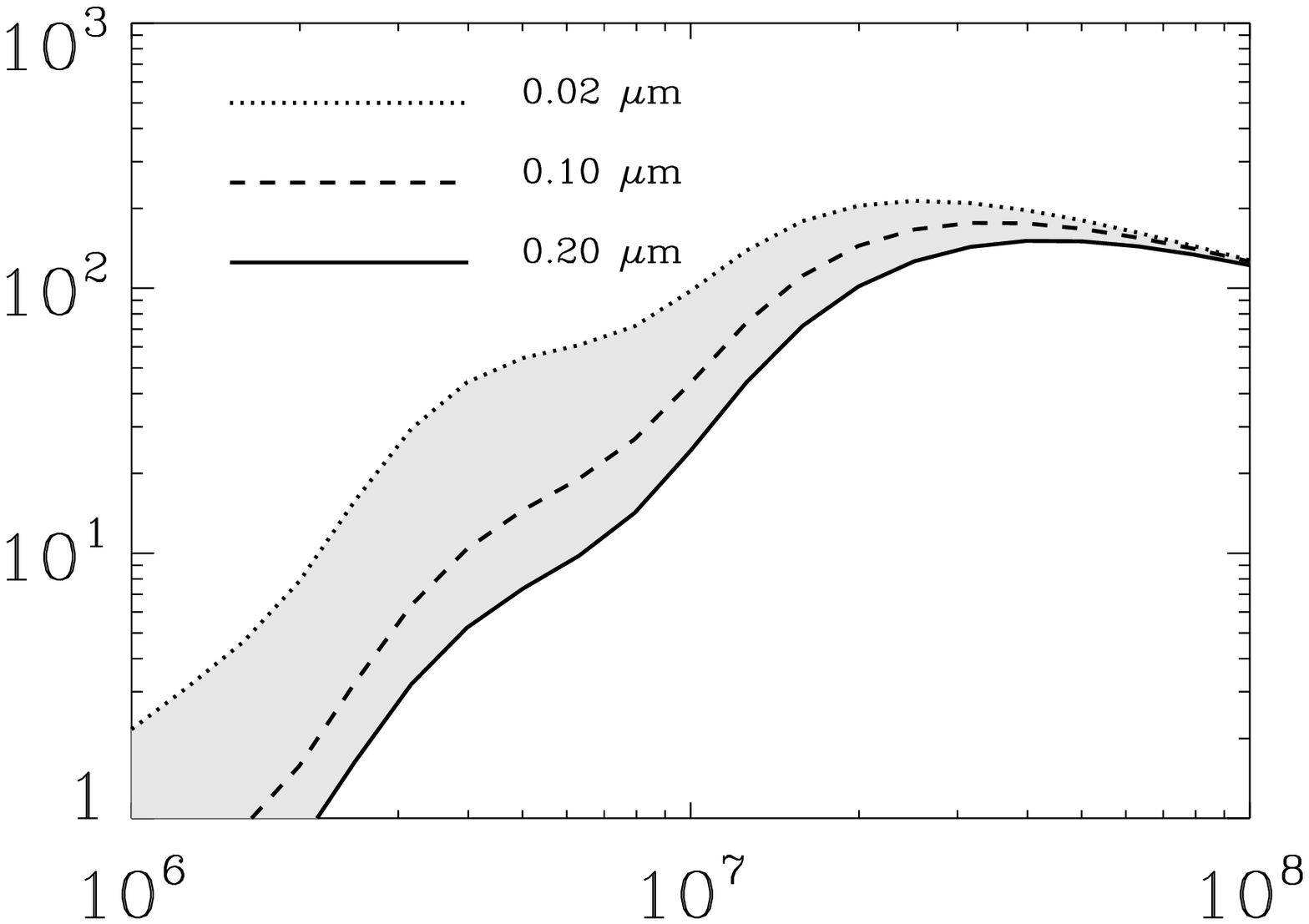} 
   \vspace{0.2in} 
   \end{tabular}
  \caption{{\footnotesize {\bf Left panel}: The cooling function of a dusty plasma via atomic processes (thick solid line) and via gas grain collisions. Calculations were performed assuming a single-sized population of 0.02, 0.1~\mic\ and 0.20~\mic\ grains (solid line) with a dust-to-gas mass ratio of 0.0062, which is the value in the local ISM of the bare silicate-graphite+PAH dust model of \cite{zubko04}. The gas cooling rate per unit volume for both processes is given by $L = n_e^2\, \Lambda (T)$. {\bf Right panel}: The value of \irx\ for single-sized dust populations with radii of 0.02, 0.1, and 0.2~\mic\ with the same dust-to-gas mass ratio as the figure on the left.}}
  \label{irx}
\end{figure}

\subsection{The Infrared to X-ray Flux Ratio}
Another important diagnostic of a dusty plasma is \irx, defined as the ratio of the IR to X-ray fluxes emitted by the gas (Dwek et al. 1987). If the dust is collisionally-heated by the gas then the total IR flux, $F_{IR}$, emitted from a gas volume $V$ is proportional to $n_e\, n_d\ \Lambda_d(T_g)\ V$, where $n_d$ is the number density of dust particles, and $\Lambda_d(T_g)$ is the cooling function (units of erg~cm$^3$~s$^{-1}$) of the gas via gas--grain collisions. The total X-ray flux, $F_X$, from the same volume is proportional to $n_e^2\ \Lambda_g(T_g)\ V$, where $\Lambda_g(T_g)$ is the cooling function of the gas via atomic processes. Thus
\begin{equation}
\label{ }
IRX \equiv \left({n_d\over n_e}\right) \, {\Lambda_d(T_g)\over \Lambda_g(T_g)}
\end{equation}
Both cooling functions represent the energy losses through collisional  processes, characterized by $\langle \sigma \, v\, E\, \rangle$ summed over all interactions in the plasma, where $\sigma$ is the cross section, $v$ is the relative velocity of colliding species, and $E$ is the energy lost in the process. 

For a given dust-to-gas mass ratio, that is, a fixed $(n_d/n_e)$ ratio, \irx\ depends only on plasma temperature. Figure \ref{irx} (left panel) shows the behavior of the atomic cooling function of a gas of solar composition as a function of gas temperature. Also shown in the figure is the gas cooling function via gas--grain collisions for a gas with a dust-to-gas mass ratio $Z_d = 0.0062$ \citep{zubko04}, and single-sized dust populations with radii of 0.02, 0.1, and 0.2~\mic. The right panel of the figure presents the value of \irx\ for the same conditions. The figure shows that for soft X-rays ($kT_e \sim 0.3$~keV, $T_e \sim 3.5\times 10^6$~K) this ratio varies between $\sim$ 3 and 20, depending on grain size. Each plasma temperature will have a different range of values, depending on the grain size distribution. Any deviation from these values will suggest that $Z_d$ is either depleted or overabundant with respect to the reference value adopted in the calculations.

\subsection{The Plasma Ionization Timescale}
The cooling rate of a plasma depends on the ionization state of its constituent ions which may not have evolved to equilibrium conditions. The ionization state of the gas is characterized by the ionization timescale, ${\cal F}$(units of cm$^{-3}$~s), defined as ${\cal F} \equiv n_e\times t_e$, where $t_e$ is the age of the shocked gas. In a fully ionized plasma with $n_e \approx n_{ion}$, ${\cal F}$ also measures the fluence of ions incident on the dust. When sputtering is the dominant grain destruction mechanism, ${\cal F}$ will directly determine the total mass of dust that is returned to the gas. 

In summary, the key parameters: plasma temperature, density, ionization timescale, and X-ray fluxes, and the dust temperature, grain size distribution, composition, and IR fluxes are closely interrelated so that knowledge of some parameters will constrain the others.
 
\section{The Evolution of the Grain Size Distribution and Dust Mass}

Consider the propagation of a shock into a dusty medium with a constant number density and a fixed dust-to-gas mass ratio $Z_d^0$. The shocked gas can be regarded as a reservoir that is continuously being filled with gas and pre-existent circumstellar dust by the expanding blast wave. If dust grains were not destroyed, the postshock gas would maintain a constant value $Z_d^0$ as the mass of shocked dust and gas evolve proportionally in time, with a functional dependence that depends on the geometry of the medium into which the blast wave is expanding. In the case where dust particles are destroyed by sputtering, the grain size distribution, and the dust-to-gas mass ratio in the shocked gas will evolve with time.

\subsection{General Equations}
A grain of radius $a_0$ that is swept up by the shock at some time $t_0$ will at time $t$ be eroded to a radius $a(t)$ given by
\begin{equation}
\label{a0a1}
a \equiv a(t)  =  a_0 + \int_{t_0}^t\, \left({da\over dt'}\right)\, dt'
\end{equation}
where
\begin{equation}
\label{dadt}
{da \over dt} \sim \sum_j\ n_j\, v_j\,  Y_j  < 0
\end{equation}
where $Y_j$ is the thermally-averaged sputtering yield of the dust by the $j$-th gas constituent. 

\begin{figure}[ht]
\center
  \includegraphics[width=4.0in]{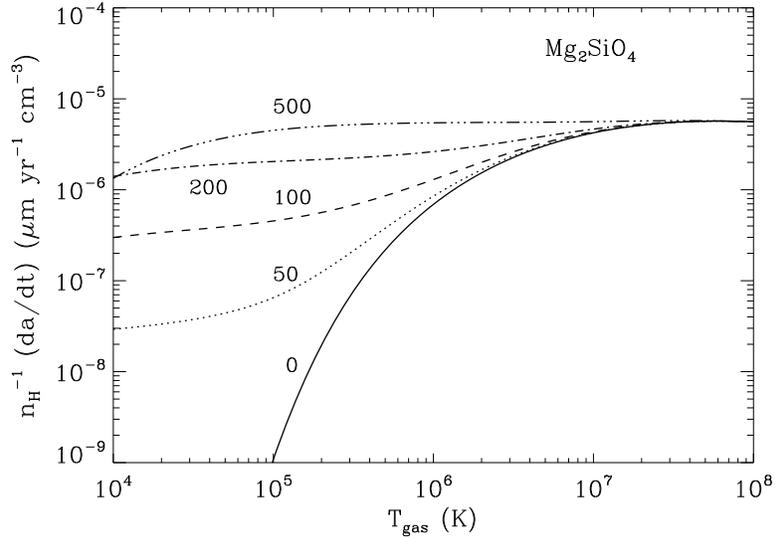} 
  \caption{{\footnotesize The absolute value of the sputtering rate (normalized to the H-number density) of silicate (Mg$_2$SiO$_4$) dust grains moving through a hot  gas of solar composition as a function  of gas temperature. The curves are marked by the grain velocity (in \kms).}}
  \label{sputrate}
\end{figure}

Dust particles swept up by a high velocity shock will move ballistically through the shock front and acquire a velocity relative to the shocked gas. The sputtering yield needs then to be averaged over a Maxwellian distribution of velocities that is displaced by  the  relative gas-grain motion from its origin  in velocity space \citep{dwek92a}. Figure \ref{sputrate} shows the temperature dependence of the sputtering rate of silicate dust grains, calculated using sputtering  yield parameters given by Nozawa et al. (2006), moving with velocity $v_{gr}$ = 0, 50, 100, 200, and 500~\kms\ through a hot gas of solar composition. In contrast to the heating of grains, their erosion by thermal and kinetic sputtering is entirely done by the ionic constituents of the gas.  For dust grains with velocities $\gtrsim 500$~\kms\ and gas temperatures above $\sim 10^6$~K the sputtering  rate is approximately constant and given by:
 \begin{equation}
\label{sput}
{da \over dt}  \approx  - 5\times 10^{-6}\, n_H({\rm cm}^{-3})\qquad \mu {\rm m}~{\rm yr}^{-1} 
\end{equation}
Equation (\ref{sput}) ignores the possible destruction of grains by evaporative grain-grain collisions that occurs in slower ($\lesssim 100$~km~s$^{-1}$) shocks by the acceleration of charged grain with MHD turbulence \citep{yan04}. 

If the shocked gas maintains a constant  composition and density then a dust grain of initial radius $a_0$ that is swept up by the shock at some time $t'$ will at time $t$ have a radius $a$ given  by:  
\begin{equation}
\label{a0a2}
a = a_0 + \left({da\over dt}\right)\, (t-t') 
\end{equation}

Equation (\ref{a0a2}) can be written in dimensionless form as:
\begin{equation}
\label{x0x}
\xi = \xi_0 -  {(t-t')\over \tau_{max}} = \xi_0-(\eta - \eta')
\end{equation}
where $\xi \equiv a/a_{max}$, $\xi_0 \equiv a_0/a_{max}$, $\eta \equiv t/\tau_{max}$, $\eta' \equiv t'/\tau_{max}$, and 
\begin{equation}
\label{ }
\tau_{max} \equiv a_{max}\, |da/dt|^{-1}
\end{equation}
is the sputtering lifetime of the largest grain in the injected size distribution, which (using eq. \ref{sput}) is numerically given by:
\begin{eqnarray}
\label{taumax}
\tau_{max} & = & 2\times 10^5\ \left[{a_{max}(\mu m)\over n_H(cm^{-3})}\right] \qquad {\rm yr} \\ \nonumber
 & = & 7.3\times 10^7\ \left[{a_{max}(\mu m)\over n_H(cm^{-3})}\right] \qquad {\rm d} 
\end{eqnarray}

Let $n_d$ be the total number density of dust grains in the preshocked gas, and $n_d(a_0)\, da_0$, the number density of grains with radii between $a_0$ and $a_0+da_0$. We assume that the grains have a size distribution in the preshocked gas given by:
\begin{equation}
\label{fsize}
n_d = \int_0^{\infty}\, n_d(a_0)\, da_0 \equiv n_d\, \int_0^{\infty}\, f(a_0)\, da_0 
\end{equation}
where $f(a_0)$ is the normalized size distribution. If the grain size distribution extends over a limited range of radii, $a_{min} \leq a_0 \leq a_{max}$, then $f(a_0) = 0$ for any $a_0 <a_{min}$ or $a_0 > a_{max}$. 

Dust grains are continuously injected into the shocked gas by the expanding SN blast wave. 
The total number of shocked grains with radii $a$ in the \{$a$, $a+da$\} interval at time $t$, $N_d(a,t)da$, is equal to the number of all dust particles of initial radius $a_0$ that were swept up at time $t'$ ($0 \leqslant t' \leqslant t$) and sputtered during the time interval $t'-t$ to radius $a$ given by eq. (\ref{a0a2}). If $\dot V(t)$ is the growth rate of the volume of the shocked gas, then $N_d(a,t)$ can be written as:
\begin{equation}
\label{nda}
N_d(a,\, t)  =   n_d\, \int_0^t \, \dot V(t')\, f(a_0)\, dt' 
\end{equation}
The lower limit of the integral, $t = 0$, corresponds to the time when the blast wave first encounters the dusty medium.

The total mass of shocked dust at any given time $t$ is given by:
\begin{equation}
\label{mdust}
M_d(t) = \int_{a_{min}-|da/dt|t}^{a_{max}}\ m_d(a)\, N_d(a,t)\, da
\end{equation}
where $m_d(a)=4 \pi \rho a^3/3$ is the mass of a dust grain of radius $a$.\\

Equation (\ref{nda}) can be written in dimensionless form:
\begin{equation}
\label{ndx}
N_d(\xi,\, \eta)  =   n_d\, \int_0^{\eta}\ \left[{dV(\eta'\,\tau_{max})\over d\eta'}\right]\, f[(\xi+\eta-\eta')\,a_{max}]\, d\eta' 
\end{equation}

This integral is a convolution of the form: $g(\eta') * f(\eta_0-\eta')$, which can be numerically evaluated for arbitrary functions using Fourier transforms.

\subsection{A Simple Analytical Solution}
An analytical solution can be derived for a pre-shocked grain size distribution given by a power law in grain radius, and a power law time dependence of $\dot V$. We write the grain size distribution as:
\begin{eqnarray}
f(a_0) & = & {\cal C}\, a_0^{-k} \qquad \qquad \   a_{min} \leq a_0 \leq a_{max} \\ \nonumber
 & = & 0 \qquad \qquad \qquad  {\rm otherwise} 
\end{eqnarray}
where ${\cal C} \equiv (k-1)/(a_{min}^{-k+1} - a_{max}^{-k+1})$ is the normalization constant. \\
The time dependence of $\dot V(t')$ can be written as:
\begin{equation}
\label{ }
\dot V(t') = \dot V_0\, \left({t' \over \tau_{max}}\right)^{\alpha}
\end{equation}
where $\dot V_0$ is a proportionality constant, and $\alpha = 2$ for a spherical blast wave expanding into a uniform interstellar medium (ISM), and $\alpha = 0$ if the blast wave expands into a one-dimensional ``finger-like'' protrusion. 

The total number density of grains in the $\{a, a+da\}$ radius interval is then given by:
\begin{equation}
\label{ndat1}
N_d(a,t)  =  n_d\, \dot V_0\, \int_0^t\, \left({t' \over \tau_{max}}\right)^{\alpha}\, f(a_0)\, dt' 
\end{equation}
Using eq. (\ref{a0a2}) to change variables from $t'$ to $a_0$, eq. (\ref{ndat1}) can be rewritten as:
\begin{equation}
\label{ndat2}
N_d(a,t)  =  \dot N_d\, \left|{da\over dt}\right|^{-1}\, {\cal C}\, \int_{a_{low}}^{a_{up}}\  \left[\left({t\over \tau_{max}} + {a \over a_{max}}\right)-\left({a_0\over a_{max}}\right)\right]^{\alpha}\ a_0^{-k}\, da_0 
\end{equation}
where $\dot N_d \equiv n_d\, \dot V_0$. 

The time dependence of $N_d(a,t)$ is contained in the limits on the integral over the grain size distribution. 
If the radius $a$ is within the range of the injected grain size distribution, that is, $a_{min} \leq a \leq a_{max}$, then $a_{low} = a$, since only grains with radii larger than $a$ could have contributed to $N_d(a,t)$. 
If the radius $a$ is smaller than $a_{min}$, then $N_d(a,t)$ is non-zero only if $a+|da/dt|t$ exceeds $a_{min}$, and $a_{low} = a_{min}$. In other words, the most recent injection of grains that could have contributed to $N_d(a,t)$ occurred at time $t-\Delta t$, where $\Delta t$ is the time required to reduce the grain radius from $a_{min}$ to $a$. The largest grains that could have been sputtered to radius $a$ during the time $t$ is equal to $a+|da/dt|t$. However, the largest grain size cannot exceed $a_{max}$, so the upper limit on the integral, $a_{up}$, is determined by the smaller of these two quantities. To summarize:
\begin{eqnarray}
\label{ }
a_{low} & = & \max \left\{a_{min},\ a\right\}\\ \nonumber
a_{up} & = &\min \left\{a_{max},\ a+\left|{da \over dt}\right|t\right\} 
\end{eqnarray}
 
For a spherical shock wave expanding into a one-dimensional protrusion ($\alpha$ = 0) the solution to eq. (\ref{ndat2}) is given by:
\begin{equation}
\label{sol0}
N_d(a,t)_{\alpha=0}  =  \dot N_d\, \left|{da\over dt}\right|^{-1}\, {{\cal C}\over (k-1)}\, \left[a_{low}^{-k+1}-a_{up}^{-k+1}\right] 
\end{equation}
At early times, when $t < a |da/dt|^{-1} \ll \tau_{max}$, $a_{up} = a$, and $a_{low} \approx a$, and the solution to eq. (\ref{sol0}) becomes:
\begin{equation}
\label{ }
N_d(a,t)_{\alpha=0}  =  \dot N_d\, t\, {\cal C}\ a^{-k} \qquad .
\end{equation}  
At late times, when $t > \tau_{max}$, $a_{up} = a_{max}$, and the asymptotic solution of (\ref{sol0}) is:
\begin{eqnarray}
N_d(a,t > \tau_{max})_{\alpha=0} & = & \dot N_d\, \left|{da\over dt}\right|^{-1}\, {{\cal C}\over (k-1)} \left[a^{-k+1}-a_{max}^{-k+1}\right] \qquad \qquad   a > a_{min}\\ \nonumber
 & = & \dot N_d\, \left|{da\over dt}\right|^{-1}\,{{\cal C}\over (k-1)}\, \left[a_{min}^{-k+1}-a_{max}^{-k+1}\right] = constant \ \ \ \  a \leq a_{min} 
\end{eqnarray}

\begin{figure}[ht]
\begin{tabular}{cc}
  \includegraphics[width=3.0in]{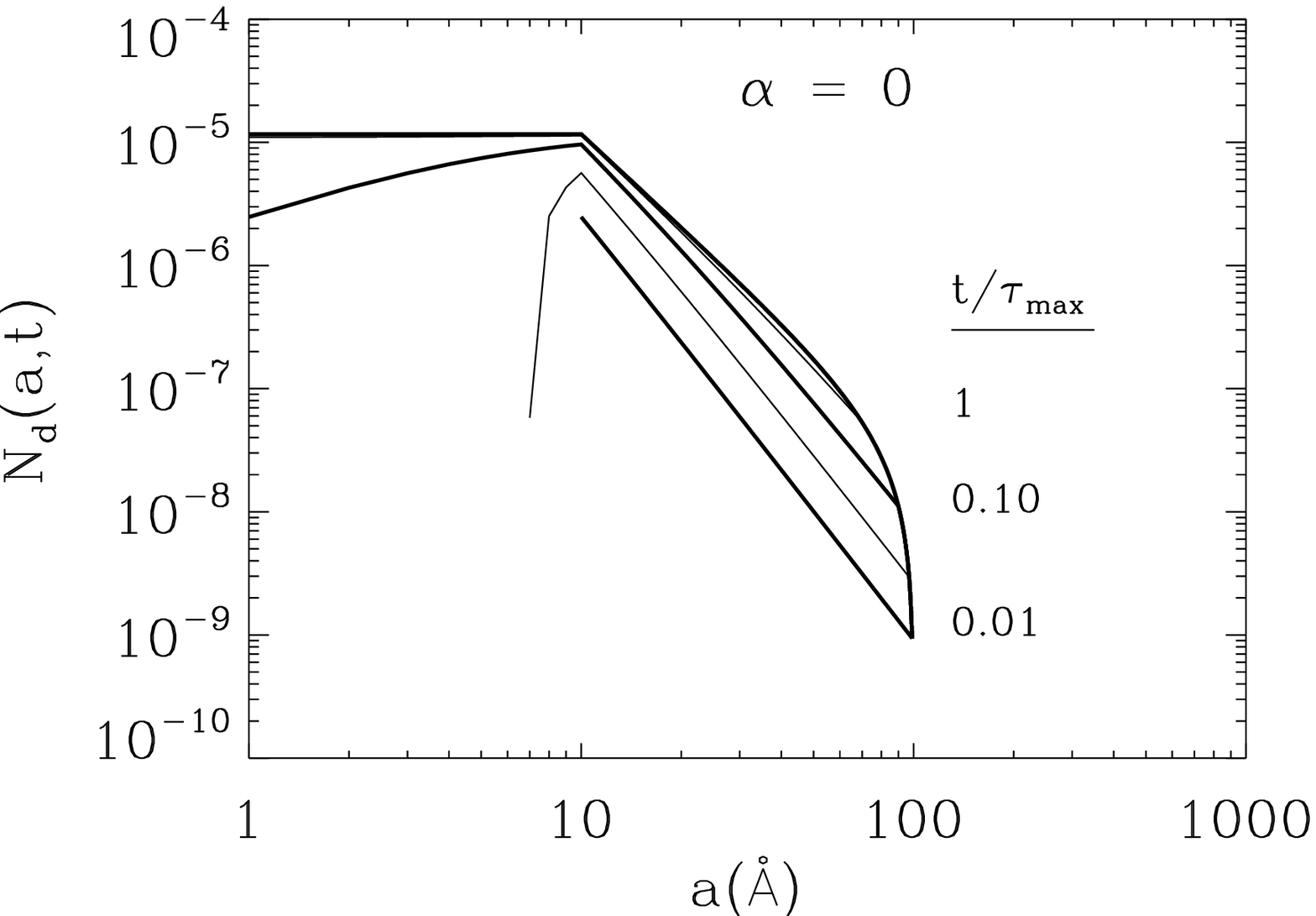} &
  \hspace{0.3in} 
   \includegraphics[width=3.0in]{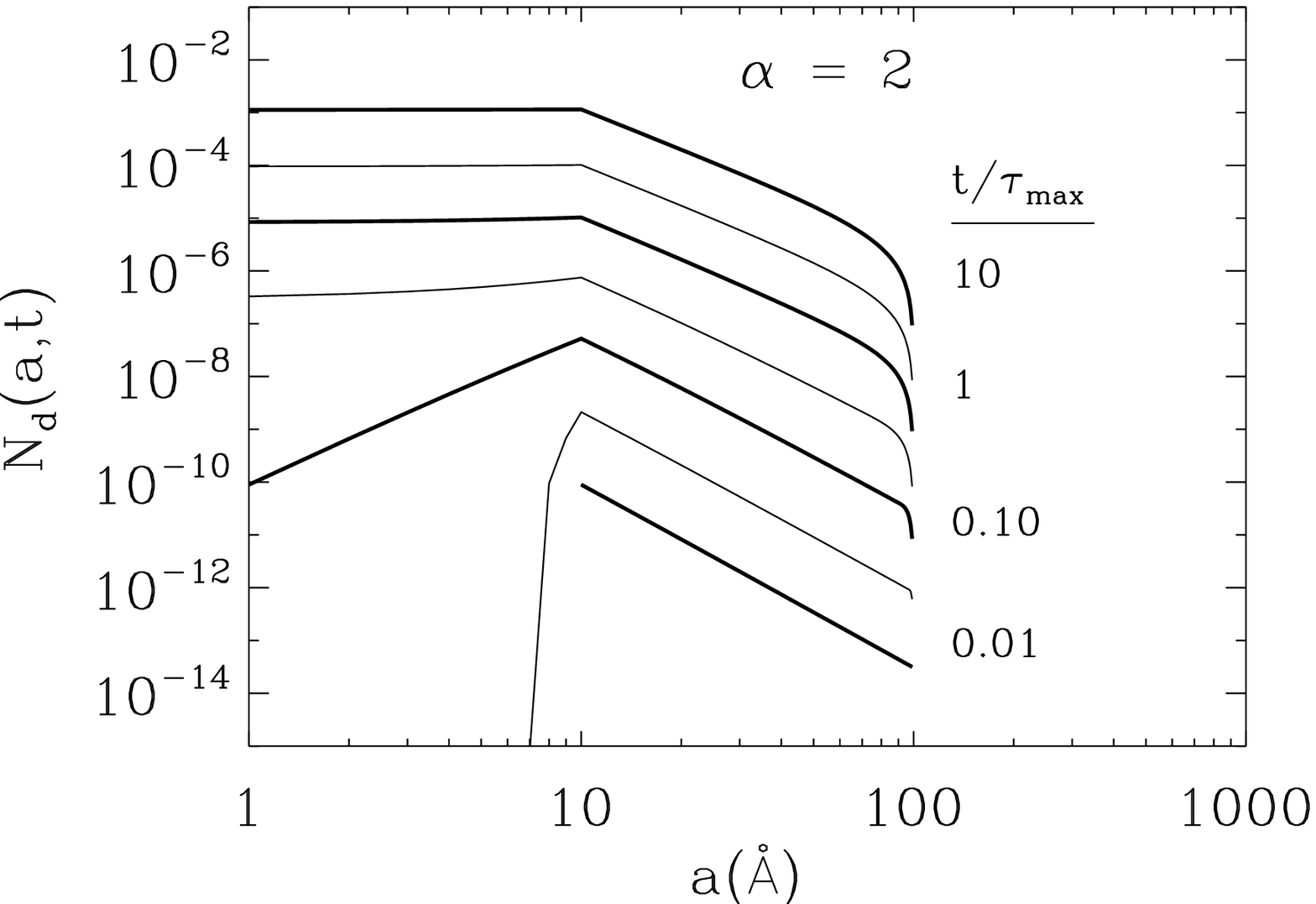} 
   \vspace{0.2in} 
   \end{tabular}
  \caption{{\footnotesize Evolution of the grain size distribution with time, measured in units of $\tau_{max}$, the sputtering lifetime of the largest grain with radius $a_{max}$ in the size distribution. Calculations were performed for a grain size distribution characterized by an $\sim a^{-3.5}$ power law in grain radii between 10 and 100~\AA. The grain destruction rate, $da/dt$, was taken to be 0.14 \AA\ d$^{-1}$, for an assumed density of 10$^4$~\cc. Bold lines are labeled by $t/\tau_{max}$. {\bf Left}: A spherical blast wave expanding into a one-dimensional protrusion ($\alpha = 0$). {\bf Right}: A spherical blast wave expanding into a uniform ISM ($\alpha = 2$).  }}
 \label{ndust02}
\end{figure}

For a shock wave expanding into a homogeneous medium ($\alpha$=2) the solution is given  by:  
\begin{eqnarray}
\label{sol2}
N_d(a,t)_{\alpha=2} & =  & \dot N_d\, \left|{da\over dt}\right|^{-1}\, {\cal C}\left\{ {1\over (k-1)}\, \left({t\over \tau_{max}} + {a \over a_{max}}\right)^2\, \left(a_{low}^{-k+1}-a_{up}^{-k+1}\right) \right.\\ \nonumber
 &  & -\  {2\over (k-2)}\, \left({t\over \tau_{max}} + {a \over a_{max}}\right)\, \left({a_{low}^{-k+2}-a_{up}^{-k+2}\over a_{max}}\right)   \\ \nonumber
 & & \left.+ \ {1\over(k-3)}\, \left({a_{low}^{-k+3}-a_{up}^{-k+3}\over a_{max}^2}\right)  \right\}
\end{eqnarray}
At late times,
when $t \gg \tau_{max}$, the first term dominates, and the asymptotic solution of eq. (\ref{sol2}) increases with time as $t^2$:
\begin{equation}
N_d(a,t > \tau_{max})_{\alpha=2}  =  \dot N_d\, \left|{da\over dt}\right|^{-1}\, {{\cal C}\over (k-1)} \left({t\over \tau_{max}}\right)^2 \ {\cal G}(a, t)
\end{equation}
where
\begin{eqnarray}
 {\cal G}(a, t) & \equiv & \left[a^{-k+1}-a_{max}^{-k+1}\right] \qquad \qquad \qquad \qquad \ \ a > a_{min}\\ \nonumber
 & \equiv &  \left[a_{min}^{-k+1}-a_{max}^{-k+1}\right] = constant \qquad \ \ \ \  a \leq a_{min} 
\end{eqnarray}

Figure \ref{ndust02} depicts the grain size distribution for different epochs. Select epochs, labeled by the dimensionless quantity $t/\tau_{max}$, are represented by bold lines. 
Calculations were performed for an initial grain size distribution characterized by an $\sim a^{-3.5}$ power law in grain radius between 10 and 100~\AA. The grain destruction rate, $|da/dt|$, was taken to be 0.14~\AA\ d$^{-1}$, calculated for the sputtering rate of silicate grains in a hot gas with a temperature and density of $\sim 10^{6-8}$~K, and 1000~\cc, respectively 

The figure illustrates the dependence of the evolution of the grain size distribution on the geometry of the ISM into which the blast wave is expanding. For a one-dimensional protrusion ($\alpha = 0$), the figure (left) shows a clear convergence of the size distribution to a fixed functional form and total number of grains for $t/\tau_{max} \gtrsim 1$. As the shock wave expands, the thickness of the shell of swept up dust increases with time. However, because of the finite grain lifetime, its thickness cannot exceed a value of $\Delta R_{sh} \approx v_{sh}\, \tau_{max}$, where $v_{sh}$ is the shock velocity. So the grain size distribution and total mass reaches a steady state limit. When the blast wave expands into a uniform medium, the shell of shocked dust reaches the same steady state thickness $\Delta R_{sh}$. However, since the surface of the shell increases as $R_{sh}^2$, where $R_{sh}$ is the radius of the blast wave, the mass of shocked gas will continue to increase. This is clearly depicted in the right panel of the figure, which shows that $N_d(a,t)$ reaches a steady state, but continues to increase with time as $t^2$.  

If grains were not sputtered in the shocked gas, then $N_d^0(a,t)\, da$, the total number of dust grains in the $\{a, a+da\}$ radius interval that are swept up by the shock at time $t$ would be:
\begin{equation}
\label{ }
N_d^0(a,t)\, da = \dot N_d\, \tau_{max}\ {{\cal C}\over (\alpha+1)}\, \left({t \over \tau_{max}}\right)^{\alpha+1} \, a^{-k}\, da
\end{equation}
Their mass, $M_d^0$, is given by eq. (\ref{mdust}) with $N_d(a,t)$ replaced by the expression above, and with $|da/dt|$ set to zero.

\begin{figure}[hb]
\begin{tabular}{cc}
  \includegraphics[width=3.0in]{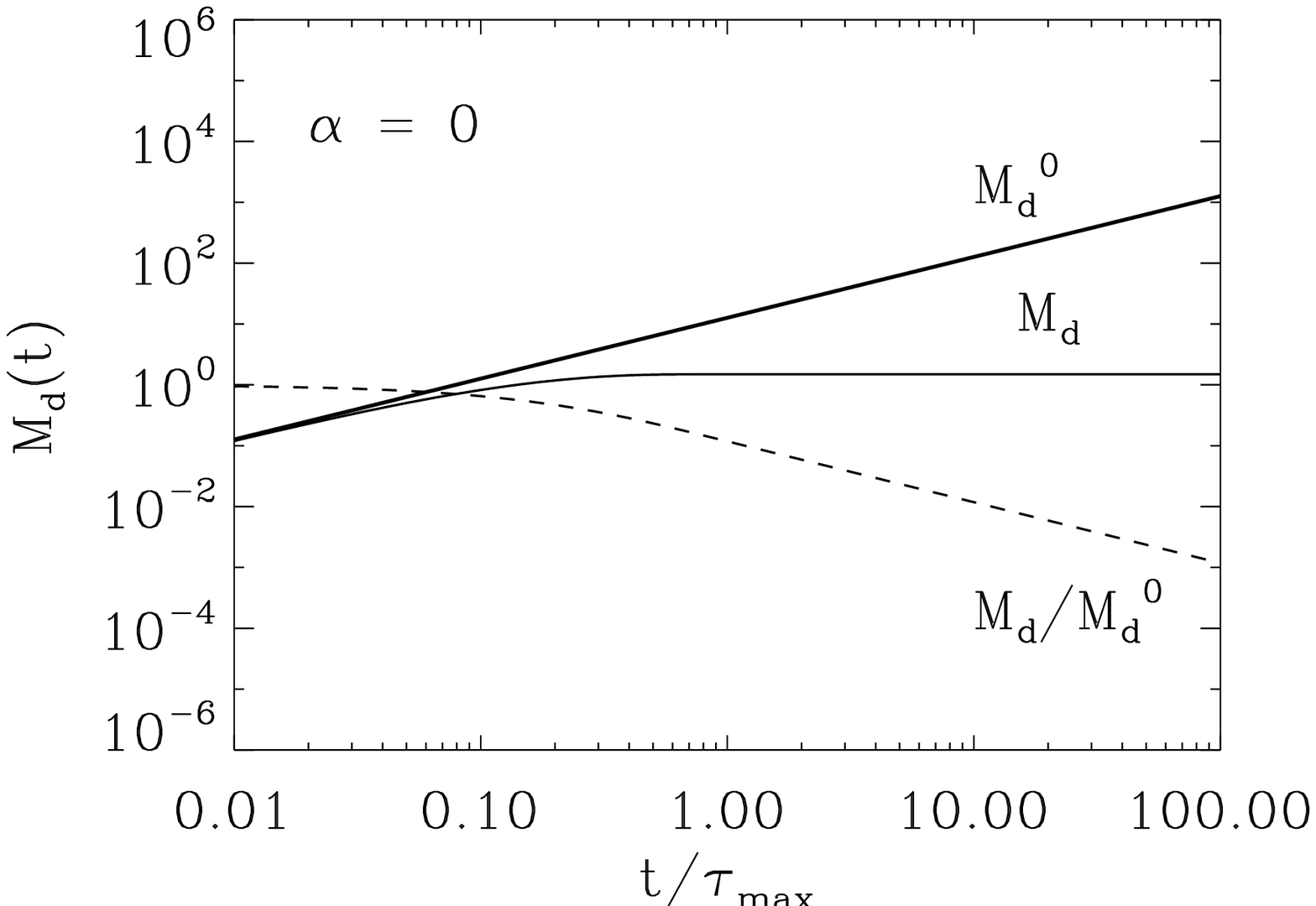} &
  \hspace{0.3in} 
   \includegraphics[width=3.0in]{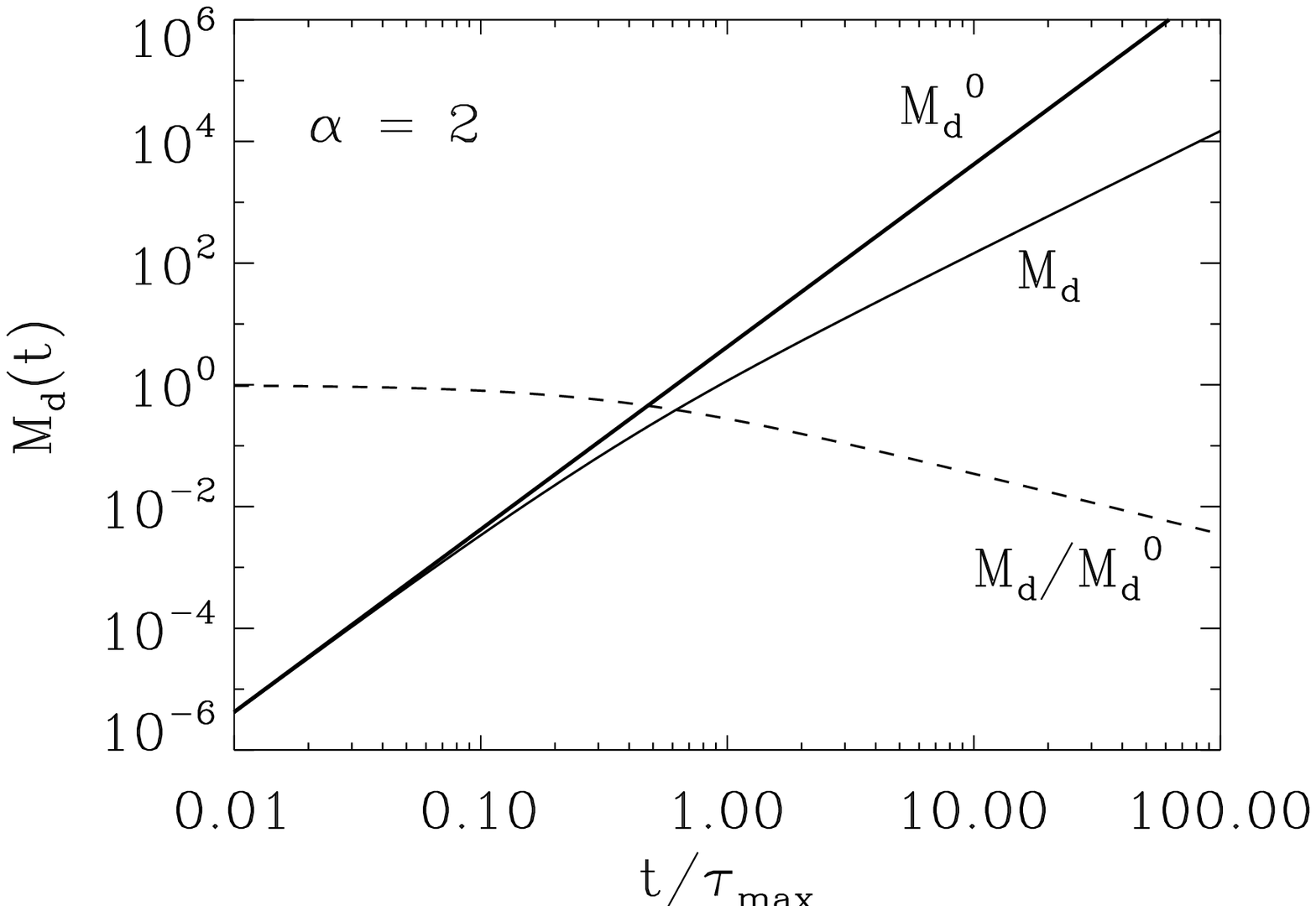} 
   \vspace{0.2in} 
   \end{tabular}
  \caption{{\footnotesize Evolution of the dust mass, $M_d$, the dust mass if grains were not sputtered, $M_d^0$, and the fraction of the surviving dust, $M_d/M_d^0$ with time, measured in units of $\tau_{max}$, the sputtering lifetime of the largest grain in the size distribution [$\tau_{max}(d) = 7.3\times 10^7\, a_{max}$(\mic)/$n_H(cm^{-3})$]. Dust and gas parameters are identical to those used in Figure \ref{ndust02}. The figure shows that the fractional change in $M_d/M_d^0$ between two epochs constrains the grain size distribution and the density of the X-ray emitting plasma. The fraction $M_d/M_d^0$ is proportional to \irx, the IR-to-X-ray flux ratio of the shocked gas. {\bf Left}: A spherical blast wave expanding into a one-dimensional protrusion ($\alpha = 0$). {\bf Right}: A spherical blast wave expanding into a uniform ISM ($\alpha = 2$). }}
  \label{mdust02}
\end{figure}

Figure \ref{mdust02} shows the evolution of dust mass for a spherical blast wave expanding into a one-dimensional protrusion ($\alpha = 0$; left), and into a uniform ISM ($\alpha = 2$; right) as a function of $t/\tau_{max}$. As explained above, when $\alpha = 0$, the mass of shocked dust, $M_d$, reaches a constant limit for $t \gg \tau_{max}$, whereas for $\alpha = 2$ the mass of the shocked dust will increase as $t^2$. If grains were not destroyed, the mass of swept up dust, $M_d^0$ would increase as $t$ for $\alpha = 0$, and as $t^3$ for $\alpha = 2$. The figure also shows the evolution of the mass fraction of surviving dust grains, $M_d/M_d^0$. This mass fraction is proportional to the dust-to-gas mass ratio in the shocked gas, and for a constant gas temperature and density, it is also proportional to \irx, the IR-to-X-ray flux ratio in the shocked gas. The figure shows that the fractional change in $M_d/M_d^0$ between two epochs constrains the value of $\tau_{max}$ given in eq. (\ref{taumax}) which in turn depends on the grain size distribution and the density of the X-ray emitting plasma. As a reminder, $\tau_{max}(d) = 7000\, a_{max}(\AA)/n_H(cm^{-3})$. For example, given a plasma density, the value of $\tau_{max}$ will depend only on $a_{max}$, the maximum grain radius. A small value of $a_{max}$ will imply a low value for $\tau_{max}$, so that large changes in $M_d/M_d^0$ occur over very short time scales. Conversely, large values of $a_{max}$ and $\tau_{max}$, will cause changes in $M_d/M_d^0$ to occur over very long time scales.

\section{{\it Spitzer} Infrared Observations of SNR~1987A}

\subsection{ The Evolution of the IR spectrum}
\begin{figure}[h!]
\begin{tabular}{cc}
  \includegraphics[width=3.2in]{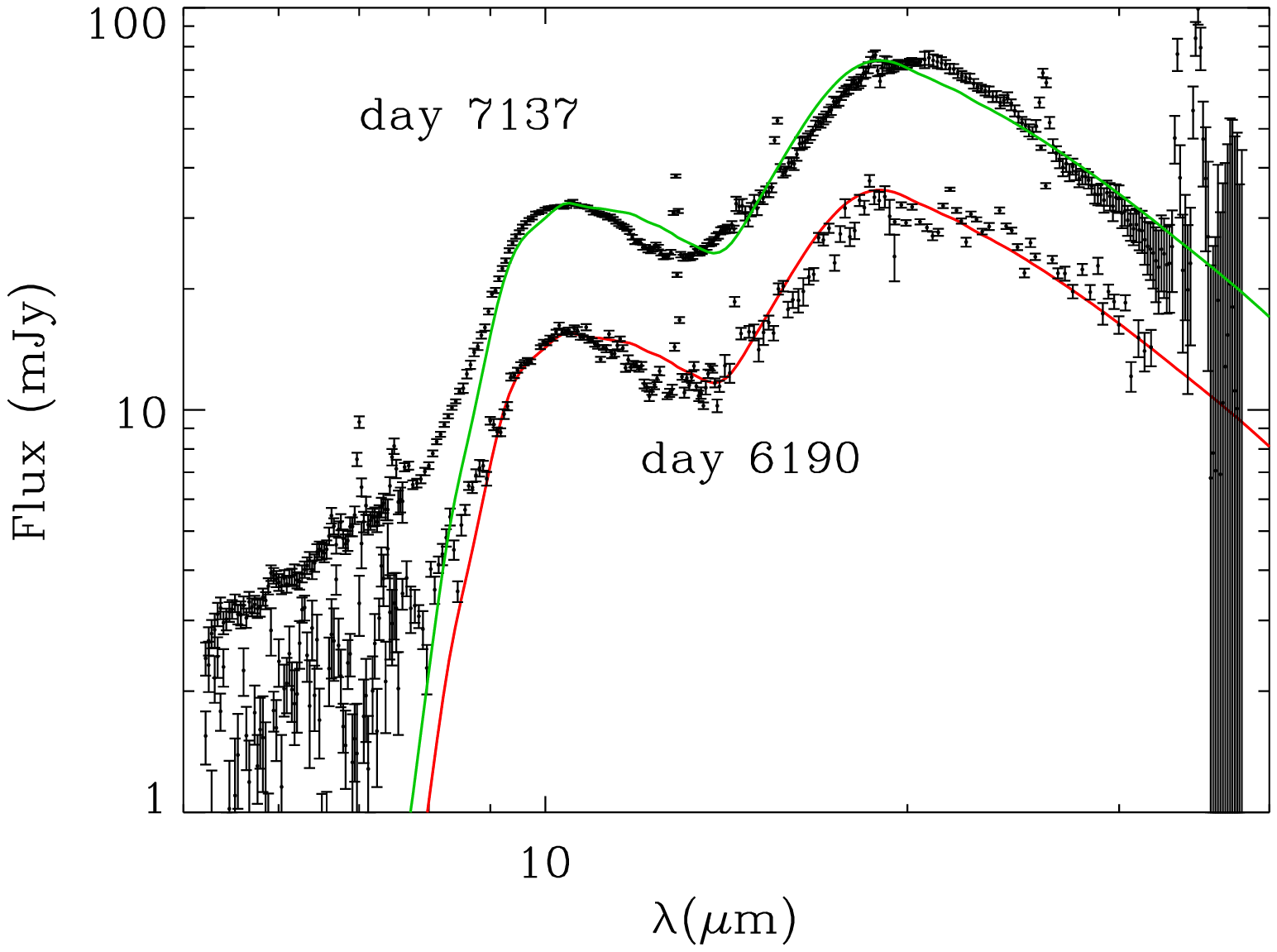} &  
   \includegraphics[width=3.2in]{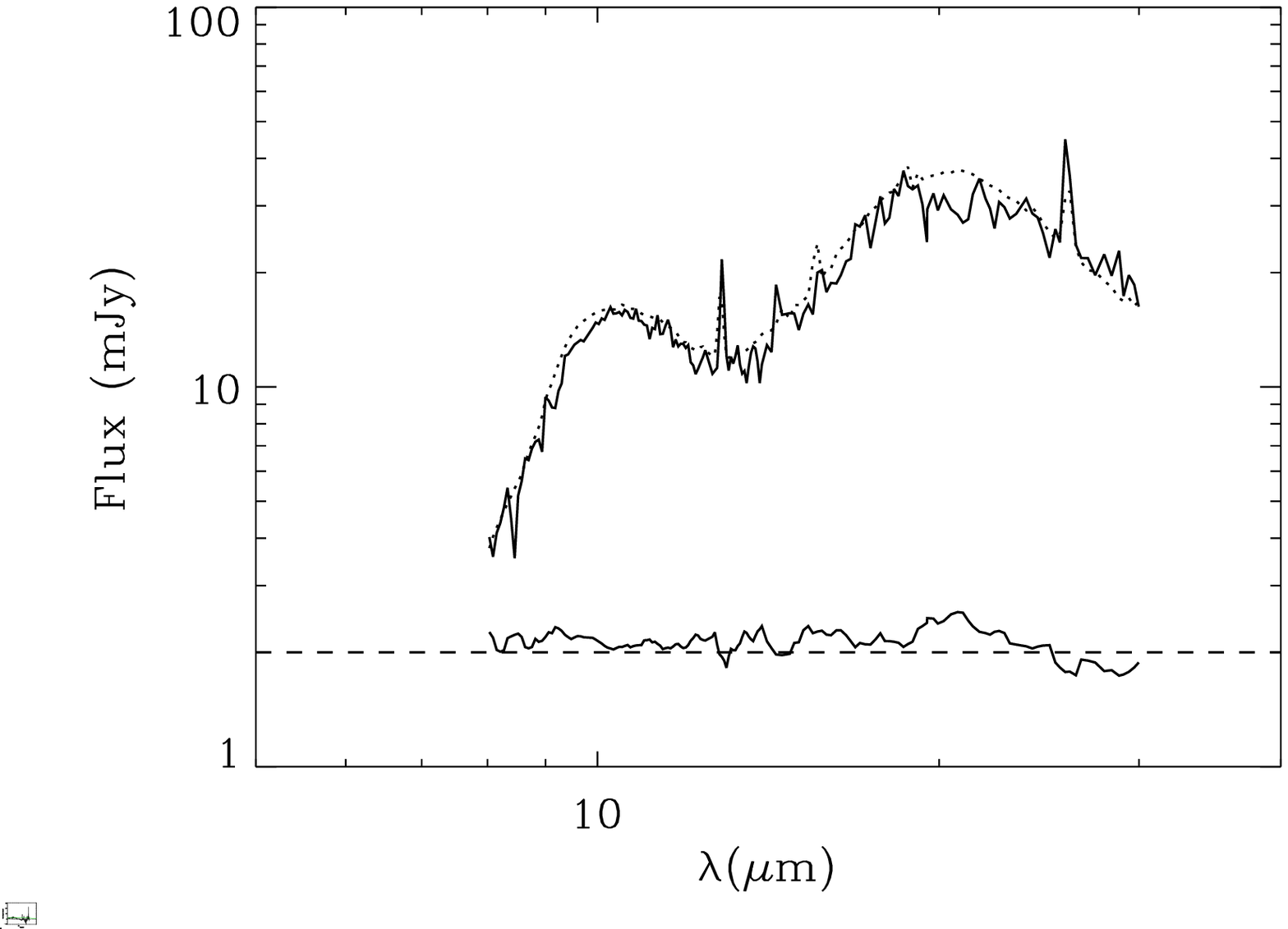}
     \end{tabular} 
  \caption{{\footnotesize {\bf Left}: The evolution of the IR spectrum of SN1987A from April 2, 2004 (day 6190 since the explosion) to September 8, 2006 (day 7137) taken with the {\it Spitzer} IRS (Bouchet et al. 2007; Arendt et al. 2007). {\bf Right}: The smoothed spectra for days 6190 (solid line) and day 7137 (dashed line), normalized to the same brightness. The lower curve shows the ratio between the two spectra, with the horizontal line being the mean value. The figure shows that the dust spectrum increased by a factor of two between the two epoch, retaining essentially an identical spectrum corresponding to silicate grains radiating at an equilibrium temperature of $180^{+20}_{-15}$~K. }}
\label{specvol}
\end{figure}

Figure \ref{specvol} shows the $ 5 - 30$~\mic\ low resolution spectra of SNR~1987A taken on February 4, 2004 (day 6190 since the explosion), and on September 8, 2006 (day 7137 since the explosion) with the Infrared Spectrograph (IRS) \citep{houck04b,houck04a} on board the {\it Spitzer} Space Telescope \citep{werner04, gehrz07}. Analysis of the spectrum taken on day 6190 revealed that the IR emission originated from $\sim 1.1\times 10^{-6}$~\msun\ of silicate grains radiating at a temperature of $\sim 180^{+20}_{-15}$~K (Bouchet et al. 2006). These circumstellar grains were formed in the quiescent outflow of the progenitor star before it exploded.
The total IR flux on day 6190 was $5.1\times 10^{-12}$~erg~cm$^{-2}$~s$^{-1}$ (Bouchet et al. 2006), and increased after 947 days (day 7137) to $10.0\times 10^{-12}$~erg~cm$^{-2}$~s$^{-1}$.
The right panel of figure \ref{specvol} presents a comparison between the two spectra, both normalized to the same 10~\mic\ intensity. The figure shows that the spectra are essentially identical, implying that the dust composition and temperature remained unchanged during the two observing periods. The lower curve in the figure is the ratio between the two spectra, emphasizing their similarity. The IR intensity increased by a factor of 2 between the two epochs.

\subsection{Determining the Grain Size Distribution}
The size distribution of collisionally-heated dust grains is constrained by the combinations of gas temperature and density that can give rise to the range of observed dust temperatures. 
The temperature of the gas giving rise to the soft X-ray component can be derived from models, and is equal to $\sim 0.3$~keV ($T_e = 3.5\times 10^6$~K; \cite{park05}), narrowing down the range of viable plasma densities and grain sizes. 

\begin{figure}[ht!]
\begin{center}
  \includegraphics[width=5.0in]{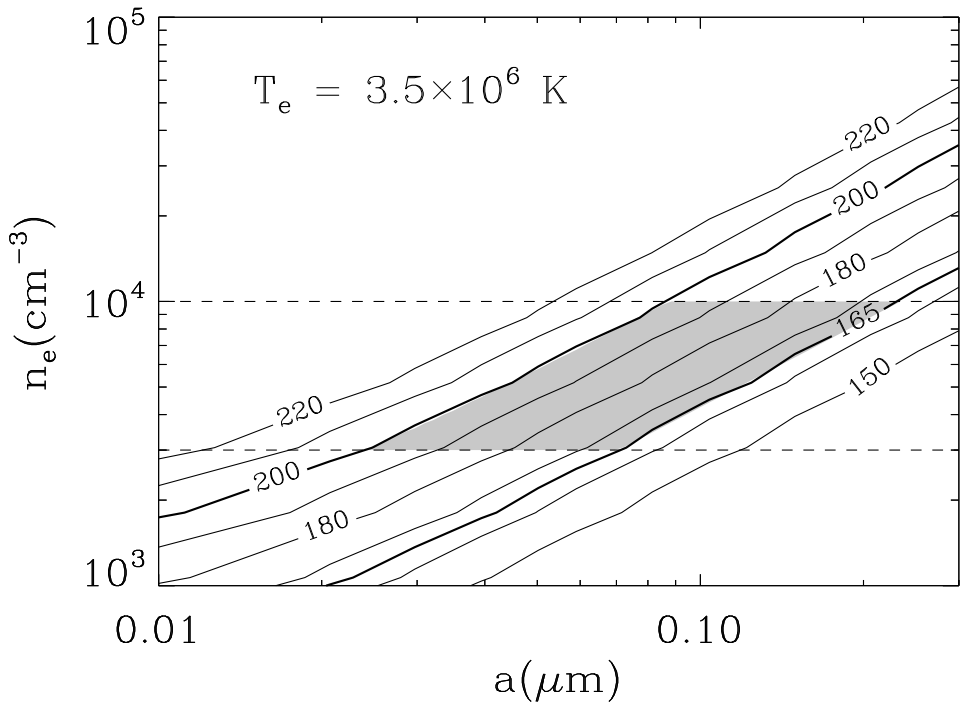}  
       \end{center}
  \caption{{\footnotesize Contours of the dust temperature as function of gas density and grain size for the given electron temperature, $T_e$, of the soft X-ray component. For the range of electron densities constrained by the plasma ionization timescale (indicated by the dashed horizontal lines), the observed range of dust temperature (indicated by the thick contours) limits the grain size distribution to be between $\sim 0.023 - 0.22$~\mic. }}
  \label{sizerange}
\end{figure}

Figure \ref{sizerange} depicts contours of the dust temperature as function of gas density and grain size for the given electron temperature, $T_e = 3.5\times 10^6$~K. 
The range of observed dust temperature falls between 165 and 200~K, and the figure shows the different combination of grain size and gas density that can give rise to this narrow range of dust temperature. 
All grain sizes are viable, provided that the gas density has the right value to heat the dust to observed range of temperatures. However, the range of viable grain sizes can be narrowed down by using the constraints on the ionization timescale of the plasma.   

The ionization timescale derived from modeling the soft X-ray spectra taken  on days 6914, 7095, and 7271 is given by ${\cal F} = n_e\, t_e \gtrsim 10^7$~\cc~d \citep{park07}. 
The ionization time, $t_e$, is constrained by the time $t_0$ when the SN blast wave first encountered the ER. Observationally, we can associate $t_0$ with the appearance of the first hot spot in the \hst\ image from April 1997, about 3700 days after the explosion \citep{pun02}. The soft X-ray light curve shows that the rise could have occurred between days 3700 and 6000. The first epoch corresponds to the first appearance of the optical knots, and the latter epoch corresponds to the time when the flux from the soft X-ray component ($kT \sim 0.3$~keV) exceeded that from the hard component ($kT \sim 2$~keV) \citep{park05}. From the mid-IR light curves \citep{bouchet06}, the energy output from the SN became ER dominated around day 4000. Adopting days 4000 to 6000 as a reasonable estimate for $t_0$ gives a range of possible ionization times of $ t_e \approx 7000 - t_0 \approx 1000 - 3000$~d. 
Using the constraints on ${\cal F}$, the corresponding limits on the electron densities are: $n_e \approx 10^4 - 3\times 10^3$~\cc.

These densities are high enough that even small dust grains with radii $\sim 10$~\AA\ will be collisionally heated to their equilibrium dust temperature. Furthermore, for a gas temperature of $T_e \approx 3.5\times 10^6$~K, an equilibrium temperature of $\sim 180$~K can only be reached at these high densities if the soft X-ray electrons are stopped in the  grains (see Figure~\ref{tdust}). The narrow range of grain temperatures then suggests that the grain size distribution should have a narrow range as well, since $T_d \sim a^{-\gamma}$ (see eq. \ref{tdust}). Figure~\ref{sizerange} shows that the constraint on the electron density, $n_e \approx (0.3-1)\times 10^4$~\cc, limits the range of viable grain sizes that can be heated up to $T_d \approx 165-200$~K to be between 0.023 and 0.22~\mic. This range is narrower than that adopted by \cite{weingartner01} to model the size distribution of LMC silicate dust (see \S4.7). 

The smaller upper limit on the size distribution of the silicates in the ER may be limited by the nucleation time scale in the outflow of the SN1987A progenitor. 
The higher lower limit on the silicate size distribution in the ER may be the result of evaporation by the initial UV flash (see \S4.3).   
\hst\ images of the ER show that it is located at a  distance of $\sim 0.7$~lyr ($6.6\times 10^{17}$~cm) from the SN. At this distance small dust particles can be evaporated by the initial UV flash that emanated from the SN \citep{fischera02}. Their calculations suggest that silicate dust particles with  radius less than $\sim 0.02$~\mic\ will be evaporated by  the flash. A population of silicate grains with a $a^{-3.5}$ power law distribution in grain radius extending from 10~\AA\ to 0.2~\mic\ will loose about 30\% of its mass.

\subsection{The Value of \irx: Constraining the Dust Abundance in the ER} 
A comparison of the IR and X-ray fluxes provides strong constraints on the dust abundance in the shocked gas. X-ray fluxes taken between days 6157 and 7271 with  the \chandra\ X-ray telescope \citep{park07} were interpolated for days 6190 and 7137 of the \spitzer\ observations. 
The total X-ray flux on day 6190, corrected for an extinction H-column density of $N_H = 2.35\times 10^{21}$~cm$^{-2}$, is $2.1\times 10^{-12}$~erg~cm$^{-2}$~s$^{-1}$, half of it radiated by the slow shock component \citep{park05}. The IR emission originates from the slow shock component which is penetrating the denser regions of the ER. This component comprises half of the observed X-ray flux. The resulting value of \irx\ on day 6190 is therefore $4.9\pm 1.1$. 

The theoretical value for \irx\ in a gas with LMC ISM abundances, taken here to be 0.6 times solar \citep{welty99a}, ranges from about 2 to 12 for soft X-rays with $T_e \sim 3\times10^6$~K. The dust abundance in the ER is therefore consistent with LMC abundances. Since the silicon abundance in the ER should not have been altered by stellar nucleosynthesis, this agreement suggests efficient condensation of silicate grains in the presupernova outflow. The dust abundance on day 6190 is a lower limit on the original pre-SN value, since some of the dust may have been evaporated by the initial UV flash from the SN.

\subsection{The Evolution of \irx}

\subsubsection{Evidence for Ongoing Grain Destruction by the SN Blast Wave}
In \S3 we consider the propagation of a shock into a dusty medium with a constant number density and a fixed dust-to-gas mass ratio. If dust grains were not destroyed, then postshock gas would maintain a constant dust-to-gas mass ratio, and the X-ray and IR fluxes from the shocked gas will evolve proportionally in time, that is, the value of \irx will remain constant. Any evolution in the value of \irx\ should therefore suggest a breakdown in the assumptions of the model.  

Observational evidence, summarized in Table 1, show that the IR flux increased by only a factor of $\sim 2$ from day 6190 to 7137.
In comparison, the extinction-corrected 0.50-2.0~keV flux increased by a factor of $\sim 3$ during the same time period to a value of $\sim 6.4\times 10^{-12}$~erg~cm$^{-2}$~s$^{-1}$ \citep{park07}. The fractional contribution of the soft X-ray component increased from 0.5 to 0.6, with no significant change in gas temperature ($kT \sim 0.3$~keV). All the increase in the soft X-ray flux can therefore be ascribed to an increase in the volume of the dense ($n_e \sim 10^4$~\cc) component of the ER that was shocked by the SN blast wave. The evolution in the X-ray and IR fluxes and the resulting value of \irx\ are summarized in Table 1. A similar evolutionary trend was reported by \cite{bouchet04, bouchet06} although absolute values of \irx\ differ from those reported here because of differences in the X-ray energy bandpasses and dust models used in the calculations.

\begin{figure}[b!]
\begin{center}
\begin{tabular}{cc}
  \includegraphics[width=3.2in]{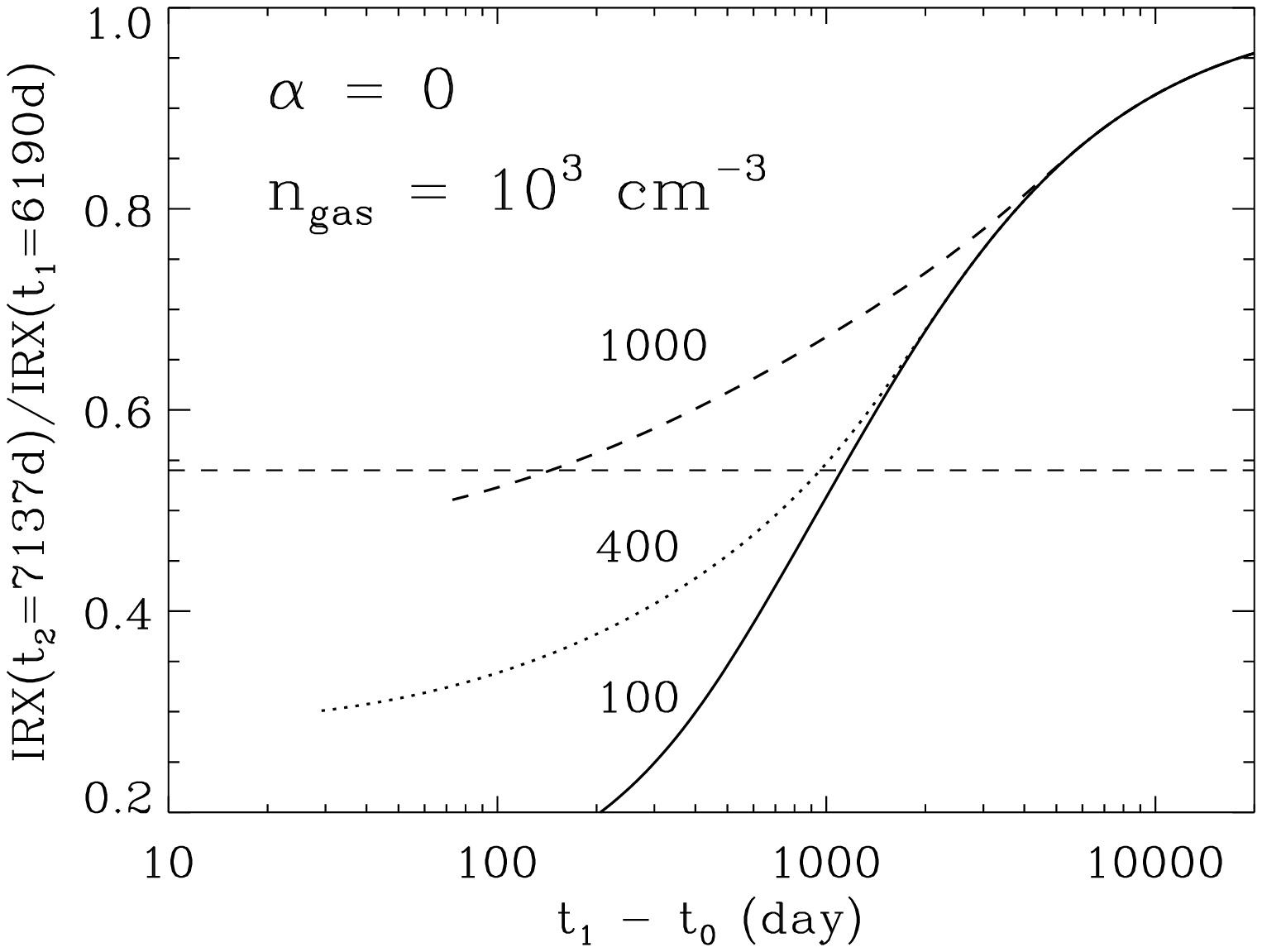} & 
   \includegraphics[width=3.2in]{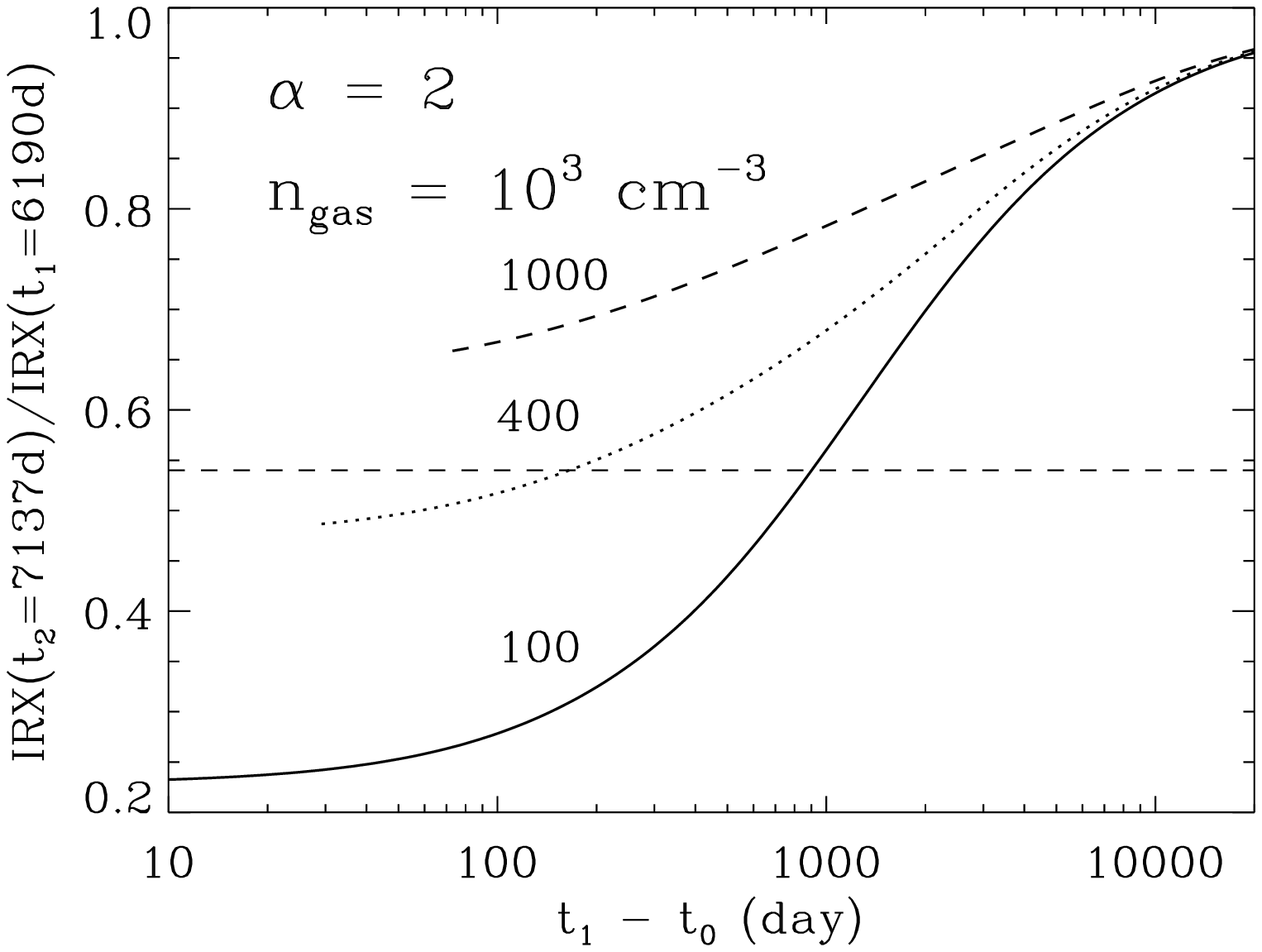}\\
     \includegraphics[width=3.2in]{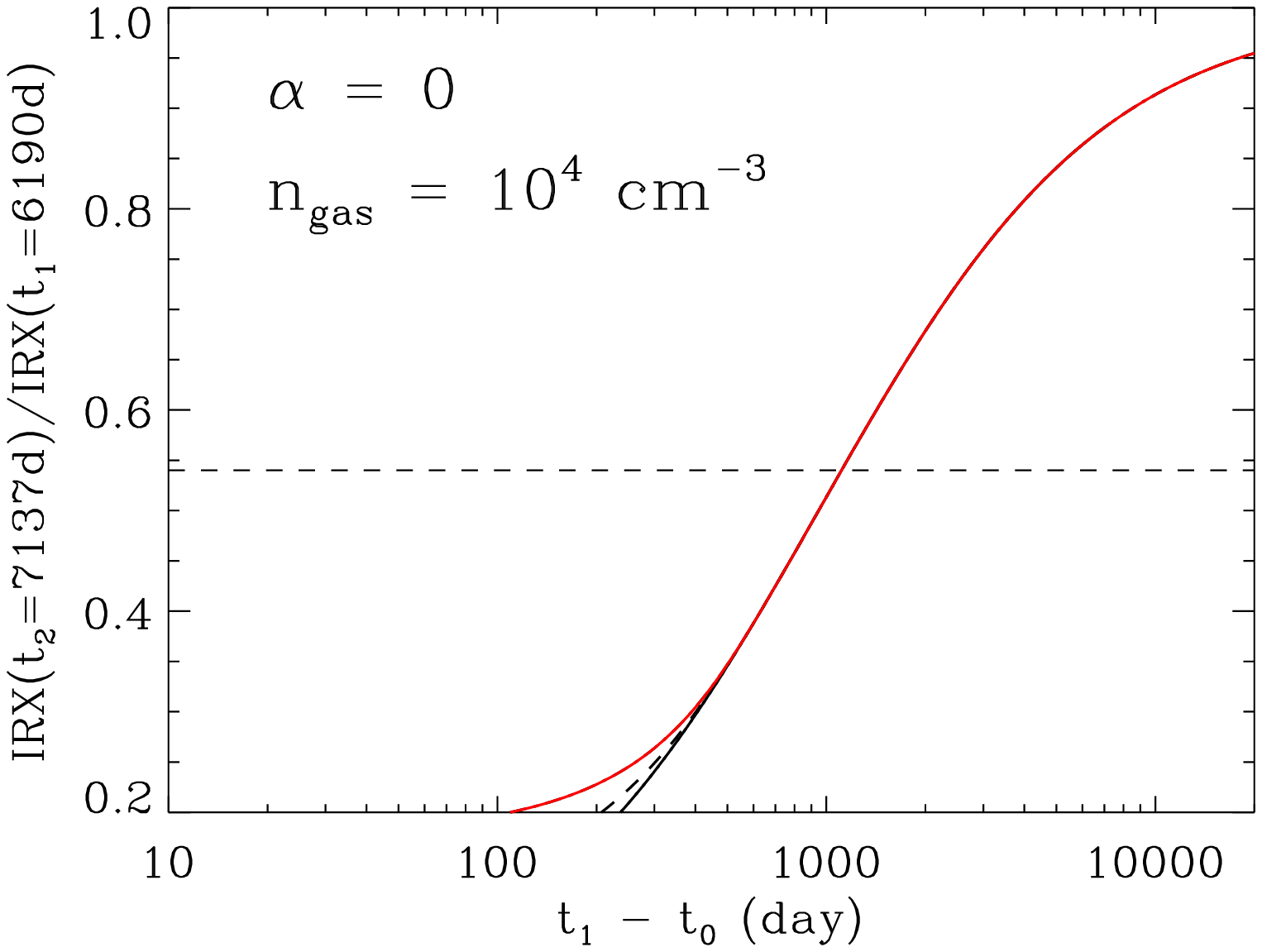} & 
   \includegraphics[width=3.2in]{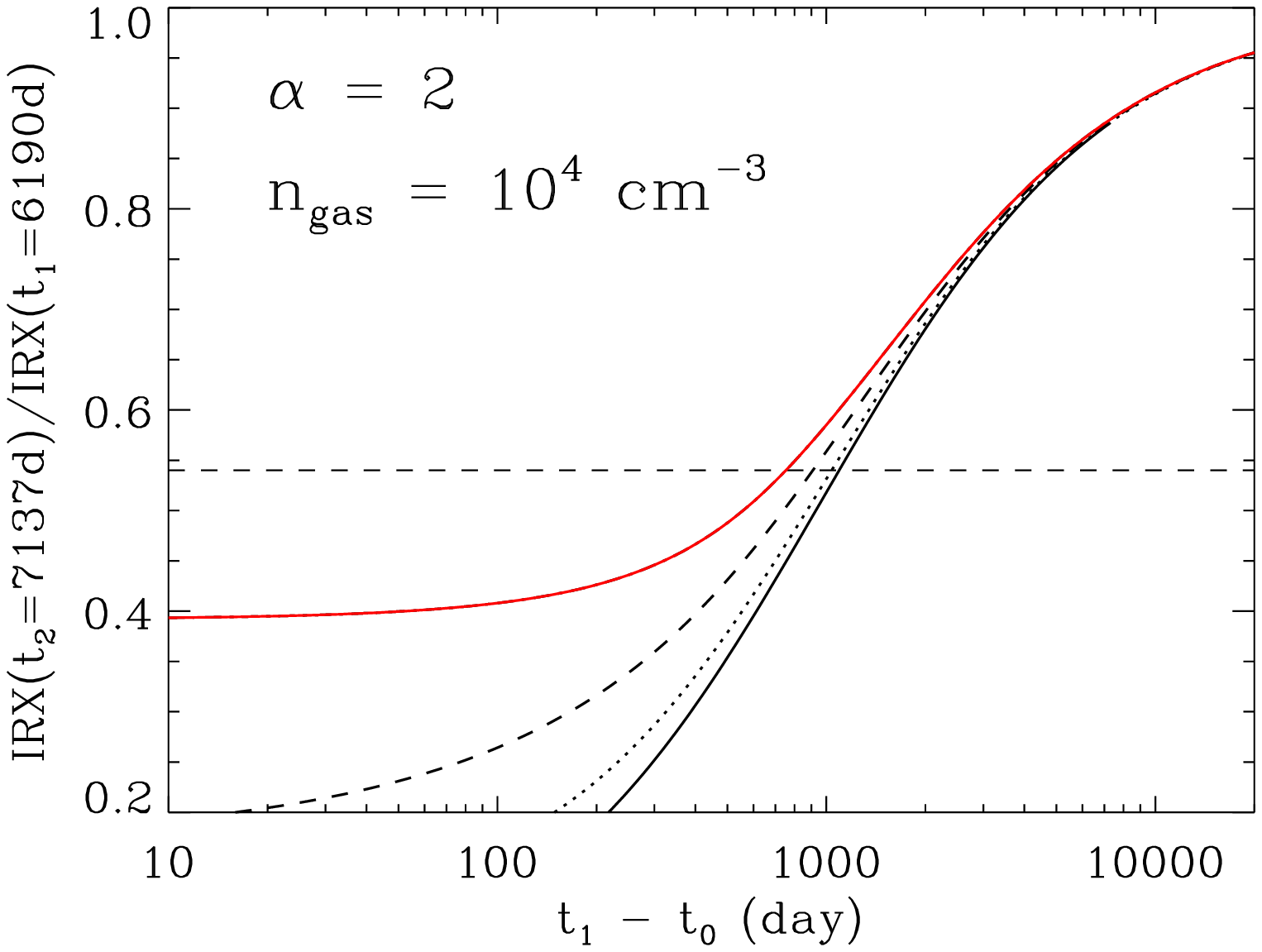} 
   \end{tabular}
       \end{center}
  \caption{{\footnotesize Evolution of the ratio \irx$(t_2)$/\irx$(t_1)$ as a function of the time difference $t_1 - t_0$, The time $t_0$ is the time since the explosion when the SN blast wave first encountered the dusty equatorial ring (ER), taken here to be the independent variable. The times $t_1$ and $t_2$ correspond, respectively, to days 6190 and 7137 since the explosion, the two epochs of near-simultaneous {\it Spitzer} and {\it Chandra} observations of the ER. The dashed horizontal line depicts the nominal value of the \irx$(t_2)$/\irx$(t_1)$ ratio which is $0.53\pm 0.16$ (see Table 1). The curves are labeled by $a_{max}$, the maximum grain size (in \AA) of the distribution. The minimum  grain size was taken  to be 10~\AA\ in all cases. The red curves correspond to a narrow grain  size distribution of the ER with $\{a_{min},\, a_{max}\}=\{0.023,\, 0.22\}$~\mic. Figures are also labeled by the value of $n_{gas}$, the density of the shocked gas. {\bf Left column}: A spherical blast wave expanding into a one-dimensional protrusion ($\alpha = 0$). {\bf Right column}: A spherical blast wave expanding into a uniform ISM ($\alpha = 2$). }}
  \label{irx02}
\end{figure}
  
If grains were not destroyed, we would expect the IR intensity to increase by a similar factor. {\it The smaller increase in the IR flux, that is, the decline in \irx, is a strong indicator that we are for the first time witnessing the actual destruction of dust in a shock on a dynamical timescale!}
If the dust composition and size distribution is uniform throughout the region of the ER that has been swept up by the shock, then \irx\ is directly proportional to the dust-to-gas mass ratio, $Z_d$, or to $M_d/M_d^0$, the ratio between the actual mass of dust in the shocked gas, and the dust mass if grains were not destroyed. The magnitude of the decrease in \irx\ between the two epochs is strongly determined by the following factors: (1) the grain size distribution in the preshocked gas of the ER; (2) the density of the X-ray emitting gas and the grain destruction efficiency, which determine the {\it rate} of grain destruction; and (3) the total time the dust is exposed to the flux of ions, which is determined by the time the SN blast wave first crashed into the ER. The latter factor determines the {\it total} mass of dust that is returned to the gas.

The dependence of the evolution of \irx\ on the grain size distribution is somewhat subtle. If the preshocked grain sizes are too large, then the fractional mass of the dust that could be destroyed during the time interval of 947 days will be too small to account for the observed decrease in the value of \irx. Conversely, if the grain sizes were too small, most of the dust mass would be destroyed, giving rise to a significantly larger than observed decrease in \irx\ between the two epochs. The right combination of grain sizes, gas density, and ion exposure time is therefore required to produce the observed dust temperature and decrease in \irx.

\subsubsection{The Grain Destruction Rate in the Hot Plasma}

Figure \ref{irx02} depicts the evolution in the $IRX(t_2)/IRX(t_1)$ ratio as a function of $t_1 - t_0$, where $t_0$ is the time, since the explosion, when the SN blast wave first crashed into the dense material of the ER, taken here to be the independent variable. The times $t_1 = 6190$~d and $t_2 = 7137$~d correspond to the two epochs of near-simultaneous {\it Spitzer} and {\it Chandra} observations of the ER. The $IRX(t_2)/IRX(t_1)$ ratio is derived by taking the the ratio of the fraction of the surviving dust mass, $M_d/M_d^0$, at the epochs $t_2-t_0$ and $t_1-t_0$ (see Figure \ref{mdust02}). Implicit in the figure is a conversion from the dimensionless time variable $t/\tau_{max}$ to absolute time.  

The observed $IRX(t_2)/IRX(t_1)$ ratio is $\sim 0.53\pm0.16$ (see Table 1), and shown as a horizontal dashed line in the figure. 
The grain size distribution used in the calculations is characterized by a $a^{-3.5}$ power law in grain radii, extending from a minimum grain size of 10~\AA\ to a value of $a_{max}$ of 0.01, 0.04, and 0.1~\mic. Results are presented for a SN blast wave expanding into a one dimensional protrusion ($\alpha = 0$, left column) and a uniform medium ($\alpha = 2$, right column) with densities of $10^3$~\cc\ (top row) and $10^4$~\cc\ (bottom row). The red line in the two bottom figures correspond to calculations performed for the grain size distribution of the ER with \{$a_{min},\, a_{max}$\} = \{0.023, 0.22\}~\mic.

The figure shows that the $IRX(t_2)/IRX(t_1)$ ratio attains its lowest value when $t_1-t_0$ is small, that is, when the first encounter of the ER with the SN blast wave occured just before $t_1$, the first epoch of {\it Spitzer} observations. Since $t_1$ is very close to $t_0$, very little grain destruction could have taken place during the $t_1-t_0$ epoch. The value of \irx$(t_1)$ is therefore close to its pre-shock value. Consequently, any subsequent destruction would lead to relatively great changes in \irx\ at $t=t_2$. Conversely, $IRX(t_2)/IRX(t_1)$  attains its largest value when  $t_0 = 0$ (a physical impossibility because of the finite time required for the SN blast wave to reach the ER). Then, the relative change in \irx\ between the two epochs will be the smallest, and $IRX(t_2)/IRX(t_1) \rightarrow 1$. To illustrate the asymptotic behavior of the ratio of \irx\ values the curves were drawn for $t_1-t_0$ values beyond the maximum physical value of 6137~d.  

Figure \ref{irx02} also shows the implicit dependence of the $IRX(t_2)/IRX(t_1)$ ratio on the grain destruction rate. When $\tau_{max}$ is small compared to the time scale of blast wave-ER interaction, the grain size distribution relaxes to its equilibrium form, and the curves of $IRX(t_2)/IRX(t_1)$ for the different grain size distributions converge to the same functional form at sufficiently large values of $t_1-t_0$. This is especially evident in the lower two figures, for which the gas density is higher, and grain destruction time scales are lower. At $t_1-t_0 \gtrsim 1000$~d the curves of $IRX(t_2)/IRX(t_1)$ for the different size distributions have all converged to the same functional form.

The figure shows that the value of $t_1-t_0$ ranges from $\approx 500 - 2000$~d, for $IRX(t_2)/IRX(t_1) = 0.53\pm 0.16$, giving values of $t_0 \approx 5700 - 4200$~d for $\alpha = 0$. The range of values for $t_0$ is consistent with the uncertainties in the time of the blast wave encounter with the ring.
This suggests that the grain size distribution derived from the X-ray constraint on the ionization timescale, the range of plasma densities, and the grain destruction efficiency used in the model are all consistent with  the observed evolution of \irx, and estimated epoch of the first interaction of the blast wave with the ER.   

The problem can also be reversed to determine the grain destruction efficiency by thermal sputtering, by adopting the epoch of $t_0 = 4000$~d and the nominal value of $IRX(t_2)/IRX(t_1) = 0.53$ as accurate descriptions of the encounter time and the rate of decrease in \irx\ in the 6190 to 7137 time interval. The the value of $\tau_{max}$ needs then to be adjusted to move the intercept from the calculated value of $t_1-t_0 \approx 1000$~d to the desired value of $\sim 2000$~d. This will require an increase in $\tau_{max}$ by a factor of 2, or a {\it decrease} in the destruction rate of the silicate grains by a factor of $\sim 2$. 
\subsection{Implications for Determining the Mass of the Circumstellar \\ Environment of SN1987A Using Light Echoes}

It is interesting to compare the dust properties derived for the ER with those derived for the progenitor's circumstellar environment from studies of the evolution and intensity of
light echoes created by the scattering of the optical light from the
supernova by the dust grains. At any given time, all points with equal
delay time lie on an ellipsoid of revolution with the SN at one focal
point and the observer at the other. Unfortunately, the ellipsoid at the
earliest epoch at which the echoes were observed was outside the ER [see Figure 11 in \cite{sugerman05b}]. As a result, the
light echoes probed only the circumstellar and interstellar media exterior to the ER.

Assuming cylindrical and reflection symmetry, \cite{sugerman05a,
sugerman05b} derived a model for the morphology of the scattering medium
consisting of: (1) a peanut-shaped contact discontinuity (CD) between the
red supergiant and main- sequence winds from the progenitor star; (2) a
structure called Napoleon's Hat (NH) constituting the waist of this
peanut; and (3) the two outer rings of the circumstellar shell (CS) that
define the
hourglass that is pinched by the ER. To model the scattered light
\cite{sugerman05a, sugerman05b} used the \cite{weingartner01} model
for interstellar LMC dust with grain radii ranging from an upper limit of 0.2-2.0~\mic\ to a lower limit of 0.00035~\mic.  By varying the relative
silicate-to-carbon dust mass ratio while maintaining an LMC
dust-to-gas mass ratio that is 0.3 times the value of the local interstellar medium, they
estimated a total nebular mass of 1.7~\msun.  They also found that the gas
density increases, the maximum grain size decreases, and the
silicate-to-carbon dust mass ratio increases as the echo samples material
that is closer to the SN. The ER, with its population of smaller pure
silicate grains, is consistent with this trend. The higher value of the minimum grain size in the ER may be the result of its proximity to the SN which caused the evaporation of grains smaller than 0.02~\mic\ by the initial UV flash. Finally, the dust abundance in the ER is consistent with that adopted by \cite{sugerman05b} for the nebula, supporting their derived value for the nebular mass.

\section{Summary}
The interaction of the SN~1987A blast wave with the complex structure of the ER has given rise to rapid evolutionary changes in the X-ray, optical and mid-IR morphology of the emission. The Gemini South mid-IR images have established that the IR emission originates from dust in the ER that is swept up by the SN blast wave, and collisionaly heated by a soft X-ray component which has a temperature of $3.5\times10^6$~K, and an ionization timescale of $n_e\, t \gtrsim 10^7$~\cc~d.
The \spitzer\ infrared observations provide important complementary information on the evolution of the interaction of the SN blast wave with the ER and the properties of the dust in the hot X-ray emitting gas. The results of our analysis are as follows:
\begin{enumerate}
\item {\it Spitzer} spectral observations on day 6190 after the explosion revealed that the dust consists of silicate dust grains radiating at an equilibrium temperature of $\sim 180\pm^{20}_{15}$~K. Subsequent observations on  day 7137 revealed that the IR flux increased by a factor of $\sim 2$, with the same dust composition and temperature remaining the same (Figure \ref{specvol}).
\item The narrow range of grain temperatures constrains the range of grain sizes and the combinations of plasma temperature and densities capable of heated the dust to the observed range of temperatures. Using the X-ray constraint on the ionization timescale we limit the grain size distribution  in the preshocked gas to be between 0.023 and 0.22~\mic. The grain size distribution may have originally extended to smaller radii, but if so, these grains were evaporated by the initial UV flash from the SN. 
\item The observed value of \irx, the IR-to-X-ray flux ratio is consistent  with that  expected from a dusty plasma with LMC abundances of heavy  elements.  
\item The value of \irx\ decreased by a factor of $\sim 0.53\pm0.16$ between days 6190 and 7137, suggesting that we are witnessing the effects of grain destruction on a dynamical timescale of the remnant.
The magnitude of the decrease in \irx\ between the two epochs is strongly determined by the following factors: the grain size distribution in the preshocked gas of the ER; the density of the X-ray emitting gas and the grain destruction efficiency; and the total time the dust is exposed to the flux of sputtering ions.
\item To follow the evolution of \irx, we constructed a model for the evolution of the grain size distribution in the shocked gas. In the model, pristine dust is continuously injected into the hot gas by the expanding SN blast wave, and destroyed by thermal  and kinetic sputtering behind the shock. The evolution of the grain size distribution resulting from the combined effect of dust injection and destruction is presented in Figure \ref{ndust02} for different geometries of the medium into which the blast wave is expanding.
\item The evolution in \irx\ represents the changes in the dust-to-gas mass ratio in the shocked gas resulting from grain destruction (see Figures \ref{mdust02} and \ref{irx02}). Given the grain size distribution and plasma density, the decrease in \irx\ between two epochs can be used to determine the epoch at which the dust was first swept up by the SN blast wave. Conversely, knowledge of the epoch at which the SN blast wave first encounters the ER can be used to determine the grain destruction efficiency in the hot gas. 
\item A self-consistent picture that emerges from the application of the model to the combined X-ray and IR observations is that of a SN blast wave expanding into a dusty finger-like protrusion of the ER with a typical LMC dust-to-gas mass ratio. The dust in the preshocked gas consists of pure silicate dust with a normal LMC dust-to-gas mass ratio and a grain size distribution limited to radii between $\sim$ 0.023 and 0.22~\mic, sufficiently large to stop the incident electrons. Smaller grain sizes may have formed in the mass outflow from the progenitor star but were probably vaporized by the initial UV flash from the SN. The SN blast wave crashed into the ER between days 4000 and 6000 after the explosion giving rise to the observed soft X-ray emission. Typical temperatures and densities of the soft X-ray emitting gas are $\sim 3\times10^6$~K and $(0.3-1)\times10^4$~\cc, consistent with those required to collisionally heat the dust to a temperature of $\sim 180\pm^{20}_{15}$~K. The plasma parameters and grain  size distribution  are consistent with the amount of grain destruction needed to account for the observed decrease in the $IRX$ flux ratio between days 6190 and 7137. At these gas densities, the onset of grain destruction  occurred  about 1200 -- 2000 days before the first \spitzer\ observations, consistent with the rise in the soft X-ray flux and the ionization time derived from X-ray models.      
\end{enumerate}

Further Gemini, ESO VLT, and {\it Spitzer} observations of SNR~1987A are in progress which, with combined X-ray observations, will shed further light on the nature of the morphology and dust properties of the circumstellar medium around the SN.

This work is based on observations made with the
{\it Spitzer} Space Telescope, which is operated by the Jet Propulsion Laboratory, California Institute of Technology, under a contract with NASA.
E.D. acknowledges partial support from HST grant GO-9114
for the Supernova INtensive
Survey (SINS: Robert Kirshner, PI), and by NASA OSS LTSA-2003-0065. 
The work of R.G.A. was supported by a grant awarded to Spitzer Cycle 3 proposal ID 30067. R.D.G. E.F.P., and C.E.W were supported by NASA through contract No. 1215746 issued by JPL/Caltech to the University of Minnesota. S.P. was supported in part by the SAO under Chandra grants GO5-6073X and GO6-7047X. 


\newpage

\begin{deluxetable}{lcccc}
\tablewidth{0pt}
\tablecaption{Observed X-ray and Infrared Fluxes From SN~1987A\tablenotemark{1}}
\tablehead{
\colhead{day number\tablenotemark{2}} &
 \colhead{ X-ray flux\tablenotemark{3}} &
  \colhead{ $f_{soft}$\tablenotemark{4}} &
 \colhead{ IR flux} &
 \colhead{ $IRX$ flux ratio\tablenotemark{5}} 
  }
 \startdata 
6190 & $(2.1\pm0.32)\, 10^{-12}$ & 0.50 & $(5.1\pm0.9)\, 10^{-12}$ & $4.9 \pm 1.1$ \nl
7137 & $(6.4\pm0.32)\, 10^{-12}$ & 0.60 & $(1.0\pm0.18)\, 10^{-11}$ & $2.6 \pm 0.5$ \nl
 \enddata
  \tablenotetext{1}{Fluxes are in units of erg~cm$^{-2}$~s$^{-1}$.}
 \tablenotetext{2}{Since the explosion.}
\tablenotetext{3}{X-ray flux in the 0.5-2.0~keV band, interpolated to the epochs of the \spitzer\ observations and corrected for an extinction column density of $N_H = 2.35\times10^{21}$~cm$^{-2}$. The error represents the uncertainty in the pile up correction factor (Park et al. 2007).}
 \tablenotetext{4}{The fraction of the 0.5-2.0~keV flux that arises from the soft ($kT \sim 0.3$~keV) X-ray component \citep{park05}.}
  \tablenotetext{5}{The ratio of  the IR to soft X-ray flux from the SN. The value of \irx\  has decreased by a factor of $0.53\pm0.16$ from day 6190 to day 7137.}
\end{deluxetable}

%
%
\end{document}